\documentclass[reprint,amsmath,amssymb,aps,prc]{revtex4-1}

\usepackage{graphicx}
\usepackage{dcolumn}
\usepackage{bm}
\usepackage{xspace}
\usepackage{dirtytalk}

\begin{document}

\title{Centrality determination with a forward detector in the RHIC Beam Energy Scan}
\author{Skipper Kagamaster}\affiliation{Lehigh University, Bethlehem PA}
\author{Rosi Reed}\affiliation{Lehigh University, Bethlehem PA}
\author{Michael Lisa}\affiliation{The Ohio State University, Columbus, OH}
\date{\today}

\begin{abstract}

Recently, Chatterjee et al~\cite{Chatterjee:2019fey} used a hadronic transport model to estimate the resolution with which various
  experimental quantities select the impact parameter of relativistic heavy ion collisions
  at collision energies relevant to the Beam Energy Scan (BES) program at the Relativistic Heavy
  Ion Collider (RHIC).
Measures based on particle multiplicity at forward rapidity were found to be significantly worse
  than those based on midrapidity multiplicity.
Using the same model, we show that a slightly more sophisticated measure 
  greatly improves the resolution based on forward rapidity particles; this
  improvement persists even when the model is filtered through a realistic simulation of
  a recent upgrade detector to the STAR experiment.
These results highlight the importance of optimizing centrality measures based on particles
  detected at forward rapidity, especially for experimental studies that search for a critical
  point in the QCD phase diagram.
Such measurements usually focus on proton multiplicity fluctuations at midrapidity, hence selecting
  events based on multiplicity at midrapidity raises the possibility of nontrivial autocorrelations.

\end{abstract}

\maketitle

\section{\label{sec:intro}Introduction}

\begin{figure}
    \centering
    \includegraphics[width=0.45\textwidth]{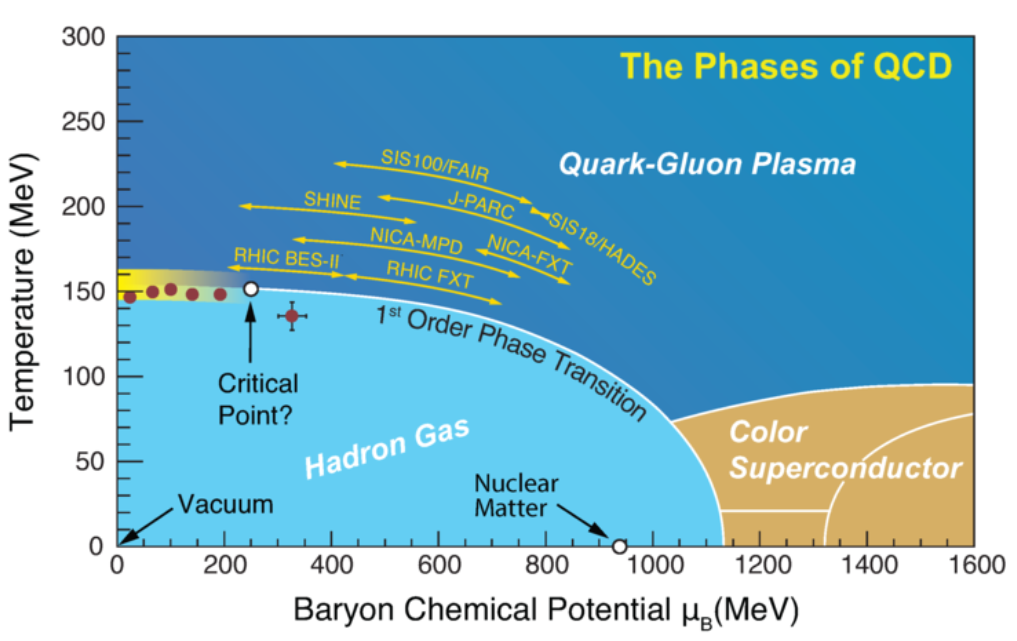}
    \caption{The QCD phase diagram, showing various states of QCD matter as functions of baryon chemical potential ($\mu_B$) and temperature (T). The CP location (and existence) is at this time not confirmed. \cite{Caines:2017search}}
    \label{fig:phase}
\end{figure}

A central goal of relativistic heavy ion collision physics is to probe the nonperturbative regime
  of quantum chromodynamics (QCD) by exploring the phase diagram of partonic matter as a function
  of thermodynamic variables.
At very high temperatures achievable at the Large Hadron Collider or Relativistic Heavy Ion
  Collider (RHIC), the quark-gluon plasma (QGP) is formed, a phase in which colored partons
  are the dynamical degrees of freedom~\cite{Adams:2005dq,Arsene:2004fa,Back:2004je,Adcox:2004mh}.
Collisions at progressively lower energies form matter with lower temperature and higher
  baryochemical potential, eventually reaching conditions at which the system is in the hadronic
  gas phase, in which partons are confined.
Understanding the nature of the transition between QGP and hadronic matter has been a
  central goal of the Beam Energy Scan (BES) program at RHIC~\cite{Stephanov:2007fk,Bowman:2008kc}.
Of particular interest is a possible critical point (CP)~\cite{Aoki:2006we,Gupta:2011wh,Fodor:2004nz,deForcrand:2002hgr,Qin:2010nq,Xin:2014ela,Shi:2014zpa,Fischer:2014ata,Lu:2015naa,Zhang:2017icm,Bazavov:2017tot,Fu:2019hdw,Fischer:2018sdj}
  at the terminus of the line characterizing a first-order phase transition on the QCD phase diagram,
  sketched in figure~\ref{fig:phase}.
 
In an infinite system in equilibrium, measured moments of conserved quantities such as charge or baryon number 
  are directly sensitive to the correlation length of the system.
A non-monotonic energy dependence of higher-order moments (e.g. the kurtosis) of the net baryon 
  distribution may reveal the a QCD critical point~\cite{Stephanov:1998dy,Bzdak:2019pkr}.
The STAR Collaboration has measured~\cite{Adamczyk:2013dal} the first four moments of the net-proton
  (a stand-in for net-baryon) distribution
  in heavy ion collisions at collision energies $\sqrt{s_{NN}}=7.7-200$~GeV, recently reporting non-monotonic
  behavior at the $3\sigma$ level~\cite{Adam:2020unf}.
This may suggest that the discovery of one of the long-sought features of bulk QCD is within reach.

However, great care must be taken in this analysis of small effects, to avoid nontrivial autocorrelations
  associated with event selection.
In particular, the protons and antiprotons used to construct the net-proton distribution are measured
  in the rapidity range $|y|<0.5$.
A critical feature of the observation is that the interesting behavior only occurs for the most
  central collisions, where centrality is estimated by the  multiplicity of charged
  particles-- {\it except for protons and antiprotons}--
  with pseudorapidity $|\eta|<1$.
In other words, one looks at (possibly correlated) fluctuations in the multiplicities of protons and antiprotons
  as a function of charged particle (except proton and antiproton) multiplicity.
Exclusion of the protons and antiprotons from the centrality measure does not trivially remove autocorrelations
  (by which we mean measurement of an event-wise quantity while selecting events with a measure directly related to that quantity),
  since very few (anti)protons are ``direct." Almost all are decay products of higher-mass baryons (e.g. $\Delta^{++}$ or $N^*$)
  whose sibling decay products are pions or other charged particles which are counted in the centrality measure.
  
A recent study by Chatterjee, et al.~\cite{Chatterjee:2019fey} based on a hadronic transport model concludes that the effect of these
  autocorrelations on the higher-order moments analysis is small.
However, the model employed in the study does not itself include critical fluctuations, so it is unclear how
  general this theoretical conclusion is.
In any event, it would be useful to check the robustness of the experimental observation by using a centrality
  estimator based on particles not emitted at midrapidity.

The STAR experiment has recently commissioned the Event Plane Detector (EPD)~\cite{Adams:2019fpo}
  to upgrade its capabilities in the BES program.
The EPD provides highly-segmented coverage for charged particles emitted in a wide pseudorapidity 
  range ($2.1<|\eta|<5.1$) and full azimuth.
An analysis that used the EPD for a centrality estimate would be free of potential autocorrelations
  discussed above.  
  
Chatterjee and collaborators~\cite{Chatterjee:2019fey} report that the summed multiplicity in the EPD
  acceptance would be a poor estimator of event centrality, because participant and spectator
  contributions are anticorrelated.
In what follows, we extend the study of Chatterjee and demonstrate that a slightly more sophisticated
  treatment of the signal in the EPD region improves the resolution substantially.
We show that this improvement persists even when the model is filtered through a realistic
  simulation of the STAR EPD, and that the detector can provide a measure of centrality
  with only slightly lower resolution than that from the STAR Time Projection Chamber (TPC, with coverage in the pseudorapidity range $|\eta| < 1$), but with no autocorrelation.
  
The paper is structured as follows: In Section~\ref{sec:methods} we describe the construction
  of global centrality estimators from collision data or the output of transport calculations.
We then discuss the EPD and limitations imposed by finite spatial and particle number resolution.
We then present an EPD-based centrality estimator that uses simple weights for the different
  regions of the EPD.
In Section~\ref{sec:results}, we compare the resolution of various centrality estimators
  both from midrapidity and forward rapidity regions.
In Section~\ref{sec:summary} we summarize and discuss the relevance these findings have for future centrality selection in the forward region of STAR.

\section{\label{sec:methods}Methods}
The Ultra-Relativistic Quantum Molecular Dynamic (UrQMD) model is a microscopic transport model used to simulate relativistic heavy ion collisions in the same energy ranges as the BES data \cite{Bass:1998ca}.
 For this study, we used 55k events per center of mass energy for Au+Au collisions at $\sqrt{s_{NN}} = $ 7.7, 11.5, 14.5, and 100k events for 19.6  GeV.
 
\subsection{Calculation of global centrality estimators}
\label{sec:GlobalVariables}
\begin{figure}
    \centering
    \includegraphics[width=0.45\textwidth]{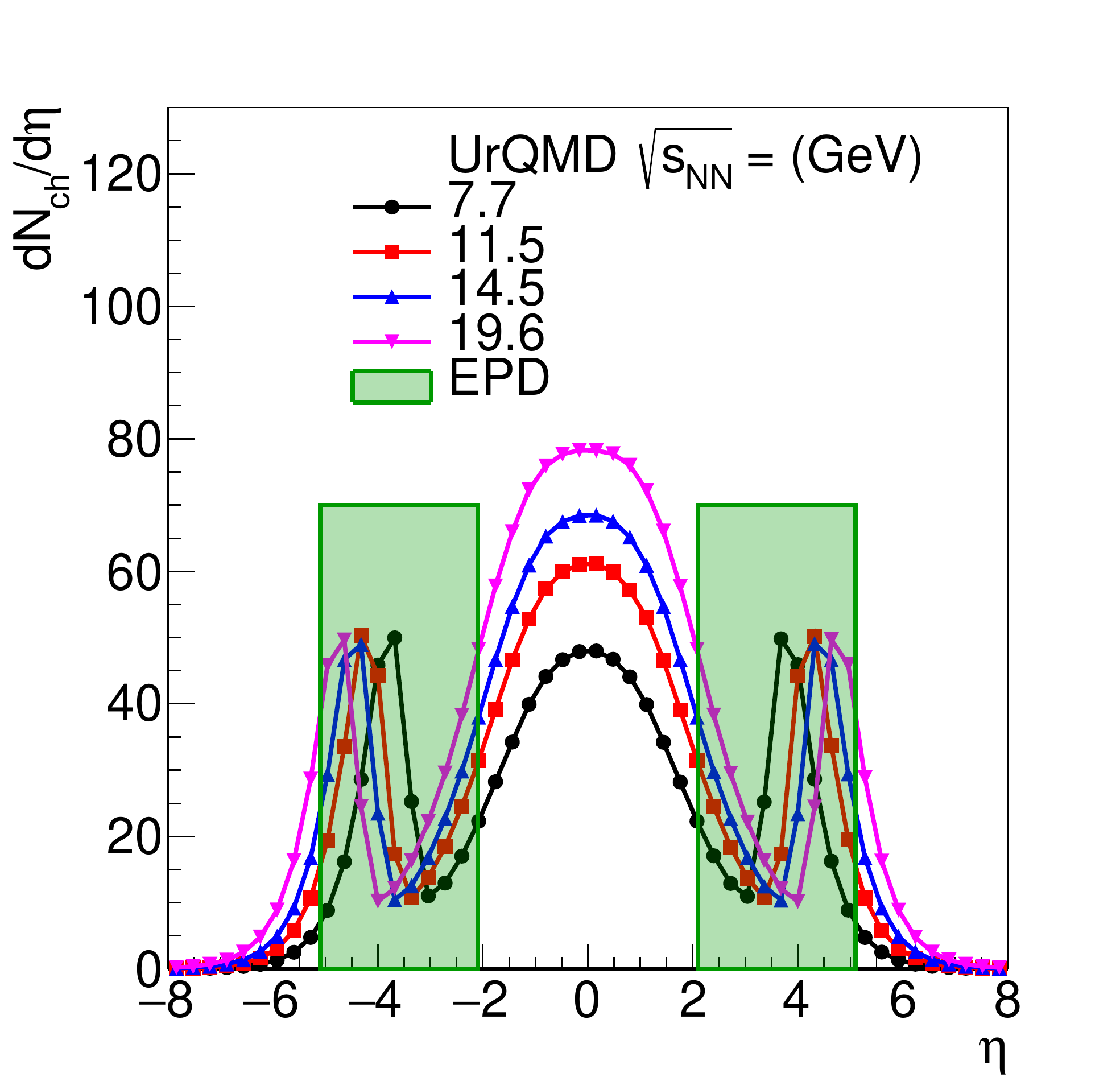}
    \caption{Charged particle pseudorapidity distributions for BES energies of $\sqrt{s_{NN}} = $ 19.6 (magenta triangle), 14.5 (blue triangle), 11.5 (red square), 7.7 (black circle) GeV.  There is no centrality selection for these distributions.  The EPD acceptance is shown as a green box.}
    \label{fig:Rapidity}
\end{figure}

Our aim is to explore the connection between experimental global observables, generically designated $X$, with the impact parameter of the collision. Naturally, this connection depends on the collision model. Glauber models~\cite{Miller:2007ri} have been successfully used to make associate particle multiplicities at midrapidity in high-energy collisions. However, at lower energies (relevant to the RHIC BES program), the spectator-participant paradigm becomes less justified. Furthermore, the longitudinal dynamics of baryon stopping, which drives the physics of the forward direction covered by upgrade detectors~\cite{Adams:2019fpo}, requires a dynamical transport model.
  
We use the UrQMD transport model~\cite{Bass:1998ca,Bleicher:1999xi}, as was used by Chatterjee~\cite{Chatterjee:2019fey}. For a given impact parameter $b$, the model begins with realistic non-smooth initial conditions and a Monte-Carlo approach; subsequent evolution of the collision is based on string dynamics and hadronic rescattering to produce final-state particles that would be measured in a detector.

In the mid-rapidity region, we compute two quantities calculated by Chatterjee for reference. The first is $X_{\rm RM1}$, defined as the charged-particle multiplicity in the pseudorapidity range $|\eta|<0.5$. The second is $X_{\rm RM3}$, the charged particle multiplicity with $|\eta|<1.0$, excluding protons and antiprotons. These centrality estimators have been used in several analyses at RHIC~\cite{Adamczyk:2015eqo,Adamczyk:2013hsi,Adamczyk:2013dal}.

\begin{figure}
    \centering
    \includegraphics[width=0.45\textwidth]{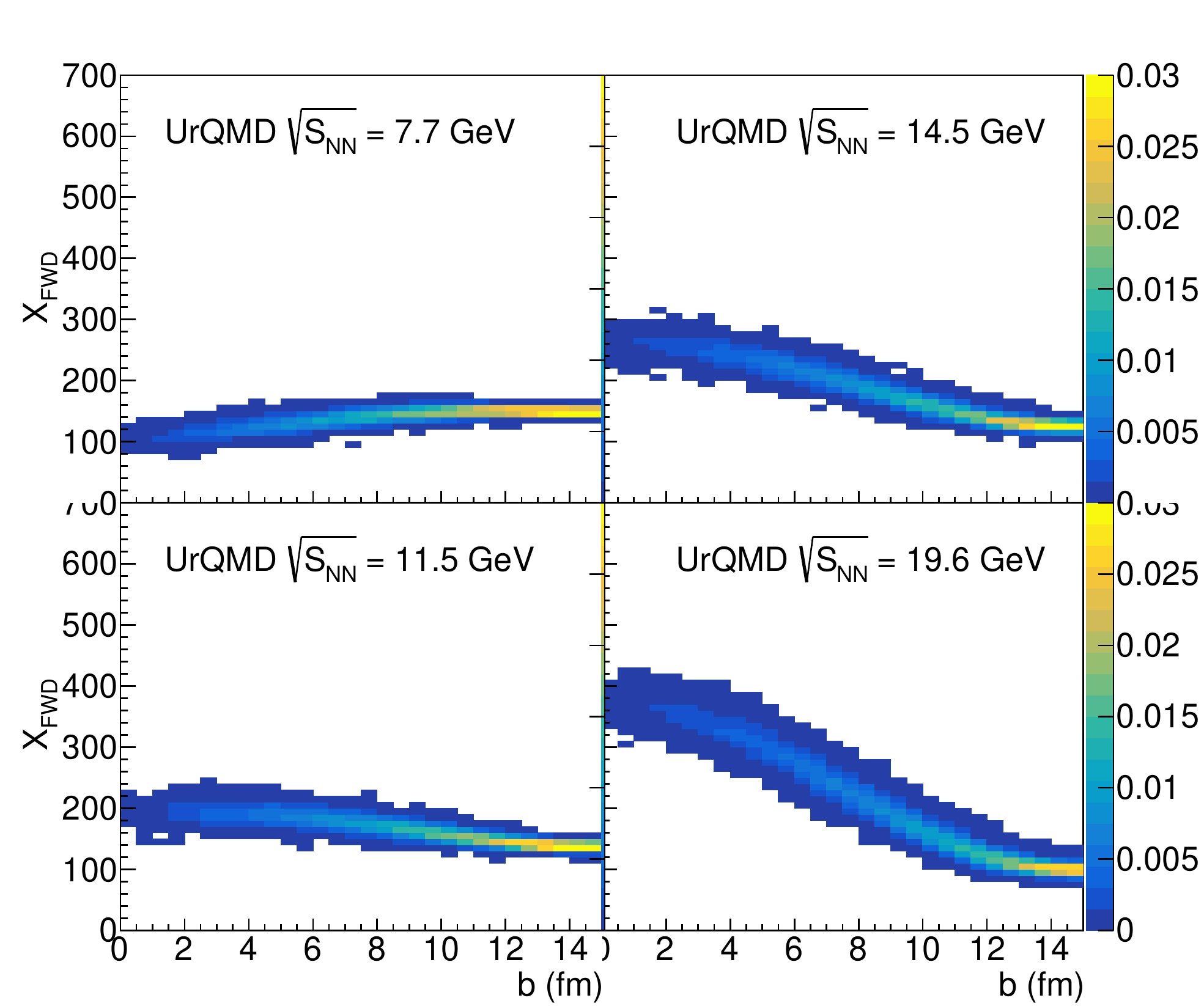}
    \caption{The sum of all charged particle yields in the EPD acceptance ($X_{FWD}$) vs impact parameter $b$ for the four collision energies considered in this paper.}
    
    \label{fig:FwdAllParticle}
\end{figure}

\begin{figure}
    \centering
    \includegraphics[width=0.45\textwidth]{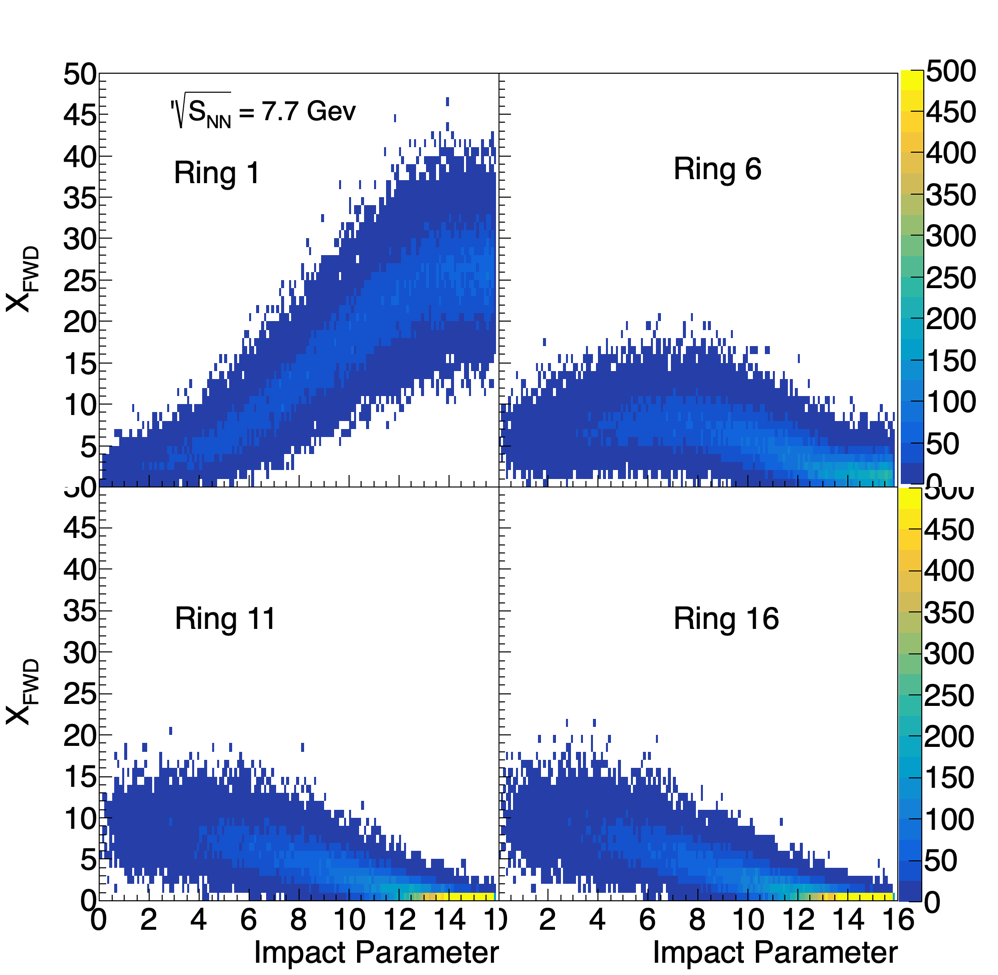}
    \caption{The sum of charged particle yields for a given EPD ring using simulated UrQMD events at $\sqrt{s_{NN}}$ = 7.7 GeV. Ring 1 is the ring closest to the beam-pipe and Ring 16 is the closest to mid-rapdity.  The precise acceptance boundaries for each ring are listed in Table \ref{table:rings}.}
    \label{fig:RingbyRing}
\end{figure}

Within the EPD acceptance, the natural analog to these measures is $X_{\rm FWD}$, the charged-particle multiplicity in the range $2.1<|\eta|<5.1$. For completeness and to compare to Chatterjee, we calculate this quantity as well, but two problems with $X_{\rm FWD}$ must be considered: it cannot be unambiguously measured in the EPD as the EPD is essentially a segmented calorimeter~\cite{Adams:2019fpo}; which we discuss this in more detail in section~\ref{sec:EPDsim}, and most importantly, the nontrivial dynamics of stopping drives yields in the forward direction. These dynamics depend on the collision energy as the spectator region ranges within the EPD acceptance (see Figure~\ref{fig:Rapidity}). They also depend on centrality, which varies the relative contributions of participants and spectators. This can be seen in Figure \ref{fig:FwdAllParticle}, which is the total charged particle yield in the EPD acceptance ($X_{FWD}$) versus the impact parameter for a variety of energies.  In this figure we see that the correlation between forward multiplicity and impact parameter decreases with decreasing collision energy due to the contribution from spectators and participants within the EPD acceptance. However, due to the segmentation of the EPD in $\eta$, it is possible to separate the different contributions, as can be seen in Figure \ref{fig:RingbyRing}.  In section~\ref{sec:RingWeights}, we present a novel approach to use this nontrivial behavior to craft an observable centrality estimator with strong correlation to impact parameter at forward rapidity.

\subsection{\label{sec:EPDsim}EPD Simulation}

The STAR TPC, which covers mid-rapidity in the experiment, measures individual charged particles with 90\% efficiency~\cite{anderson2003star}, so a centrality measure based on multiplicity is appropriate. This is not the case for the STAR EPD, which consists of 744 tiles of 1.2-cm-thick scintillator~\cite{Adams:2019fpo}. When a relativistic charged particle passes through a tile, it deposits a small amount of energy, $dE_1$, probabilistically according to the Landau distribution $\rho(dE_1)$. This distribution is characterized by two parameters: the most probable value ($dE_{\rm MPV}$) for energy loss and a width ($dE_{\rm WID}$); both the energy scale and the relative width ($dE_{\rm WID}/dE_{\rm MPV}$) are determined by the thickness of the scintillator. For the EPD, $dE_{\rm WID}/dE_{\rm MPV}\approx0.2$~\cite{Adams:2019fpo}.

\begin{table}
\centering
\begin{tabular}{|c|c|c||c|c|c|} 
 \hline
 & & & & & \\ 
 \hspace{0.1cm}EPD Ring\hspace{0.1cm} & $|\eta_l|$ & $|\eta_h|$ & \hspace{0.1cm}EPD Ring\hspace{0.1cm} & $|\eta_l|$ & $|\eta_h|$ \\ [1.5ex] 
 \hline\hline
  & & & & & \\
 1 & \hspace{0.1cm}5.09\hspace{0.1cm} & \hspace{0.1cm}4.42\hspace{0.1cm} & 9 & \hspace{0.1cm}2.81\hspace{0.1cm} & \hspace{0.1cm}2.69\hspace{0.1cm} \\
 2 & \hspace{0.1cm}4.42\hspace{0.1cm} & \hspace{0.1cm}4.03\hspace{0.1cm} & 10 & \hspace{0.1cm}2.69\hspace{0.1cm} & \hspace{0.1cm}2.59\hspace{0.1cm} \\
 3 & \hspace{0.1cm}4.03\hspace{0.1cm} & \hspace{0.1cm}3.74\hspace{0.1cm} & 11 & \hspace{0.1cm}2.59\hspace{0.1cm} & \hspace{0.1cm}2.50\hspace{0.1cm} \\
 4 & \hspace{0.1cm}3.74\hspace{0.1cm} & \hspace{0.1cm}3.47\hspace{0.1cm} & 12 & \hspace{0.1cm}2.50\hspace{0.1cm} & \hspace{0.1cm}2.41\hspace{0.1cm} \\
 5 & \hspace{0.1cm}3.47\hspace{0.1cm} & \hspace{0.1cm}3.26\hspace{0.1cm} & 13 & \hspace{0.1cm}2.41\hspace{0.1cm} & \hspace{0.1cm}2.34\hspace{0.1cm} \\
 6 & \hspace{0.1cm}3.26\hspace{0.1cm} & \hspace{0.1cm}3.08\hspace{0.1cm} & 14 & \hspace{0.1cm}2.34\hspace{0.1cm} & \hspace{0.1cm}2.27\hspace{0.1cm} \\
 7 & \hspace{0.1cm}3.08\hspace{0.1cm} & \hspace{0.1cm}2.94\hspace{0.1cm} & 15 & \hspace{0.1cm}2.27\hspace{0.1cm} & \hspace{0.1cm}2.20\hspace{0.1cm} \\
 8 & \hspace{0.1cm}2.94\hspace{0.1cm} & \hspace{0.1cm}2.81\hspace{0.1cm} & 16 & \hspace{0.1cm}2.20\hspace{0.1cm} & \hspace{0.1cm}2.14\hspace{0.1cm} \\ [1ex] 
 \hline
\end{tabular}
\caption{List of the low ($|\eta_l|$) and high ($|\eta_h|$) values for the $|\eta|$ ranges of each EPD ring \cite{Adams:2019fpo}.}
\label{table:rings}
\end{table}

\begin{figure}
\includegraphics[width=0.45\textwidth]{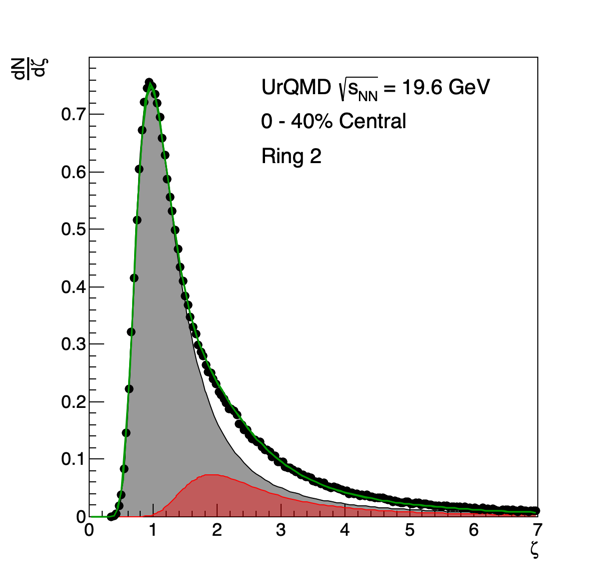}
\caption{
A low-flux tile which is dominated by either one or two particles passing per collision events. The coloured peaks show the 1 (grey) and 2 (red) particle Landau distributions, whereas the spectrum (black circles) is a sum of the convoluted distributions (green line).
}
\label{fig:zSpectraLowFlux}
\end{figure}

As is the practice experimentally, we will quantify the signal in a tile by normalizing to the peak in the single-particle distribution:
 \begin{equation}
     \zeta \equiv \frac{dE}{dE_{\rm MPV}}  .
 \end{equation}
The single-particle distribution $\frac{dP_1}{d\zeta}$ then peaks at unity and has Landau width $\sim0.2$. Figure~\ref{fig:zSpectraLowFlux} shows the simulated distribution for a tile that is dominated by one or two particles crossing per collision event.

\begin{figure}
\includegraphics[width=0.45\textwidth]{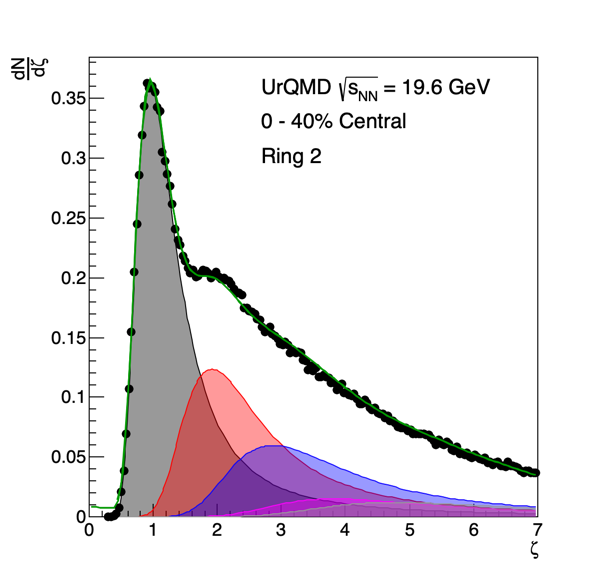}
\caption{
A high-flux tile which experiences a range of particle multiplicities, resulting in a spectrum (black circles) that is a sum of convoluted Landau distributions (green line). The coloured peaks show the 1 (grey), 2 (red), 3 (blue), 4 (purple) and 5 (dark grey) particle Landau distributions.}
\label{fig:zSpectraHighFlux}
\end{figure}

Depending on the collision energy and centrality, $N>1$ charged particles may pass through a tile in a given collision event, each leaving energy independently. The total energy in such events follows a distribution:
\begin{equation}
    \frac{dP_{N}}{d\zeta}\left(\zeta\right) = \int^{\infty}_{0}d\zeta_1 \frac{dP_1}{d\zeta}\left(\zeta_1\right)\frac{dP_{N-1}}{d\zeta}\left(\zeta-\zeta_1\right) .
\end{equation}
The total energy deposited into each tile is recorded; for every collision event, the distribution is a weighted sum of $\frac{dP_{N}}{d\zeta}$, seen for $N\leq5$ as the histogram in Figure~\ref{fig:zSpectraHighFlux} (distributions for $N=1,2,3,4,5$ are shown as shaded curves). This distribution has been shown to reproduce the EPD experimental measurements closely~\cite{Adams:2019fpo}.

In general, the more particles that pass through a tile in a collision event, the larger the measured $\zeta$ will be.
It can seem natural to construct EPD-based centrality measures analogous to those used in the TPC by substituting $\zeta$ for multiplicity. However, the most probable value for $\frac{dP_{N}}{d\zeta}$ is not found at $\zeta=N$, and there is clearly considerable overlap between $\frac{dP_{N}}{d\zeta}$ distributions. Indeed, for a low-flux tile (c.f. Figure~\ref{fig:zSpectraLowFlux}), it is most accurate to assume $N=1$ in any collision event for which $\zeta>0$. Assuming $N=\zeta$ (which is not even correct ``on average'' as the Landau distribution does not have a well-defined mean) in this case only builds in unwanted noise for an event-wise centrality measure. Even in the high-flux tile of Figure~\ref{fig:zSpectraHighFlux}, a $\zeta=4$ event is most likely caused by three particles having passed through the tile.

How these effects influence the correlation between a centrality estimator and the true impact parameter depends on the collision model and the detector geometry. Using the UrQMD model, the collisions are assumed to take place at the center of the STAR experiment (for simplicity) and, upon emission, charged particles are assumed to propagate in a straight line until they strike the EPD. A precise geometric model of the active elements of the EPD~\cite{Adams:2019fpo} is used to register the passage of charged particles through each tile. Each charged particle deposits energy according to the Landau distribution, as discussed above, with the net signal from a tile being the sum of all deposited energy.\footnote{Our simulation does not include the effects of the magnetic field in the STAR experiment; however, at these rapidities, this is relatively unimportant for measuring multiplicity as a function of $\eta$. Secondary scattering and production in surrounding material (magnet iron, beam pipe, etc) is a larger effect that is not accounted for, in our simulation. Exploring corrections due to these effects is outside the scope of the present study.}

\begin{figure}
    \centering
    \includegraphics[width=0.45\textwidth]{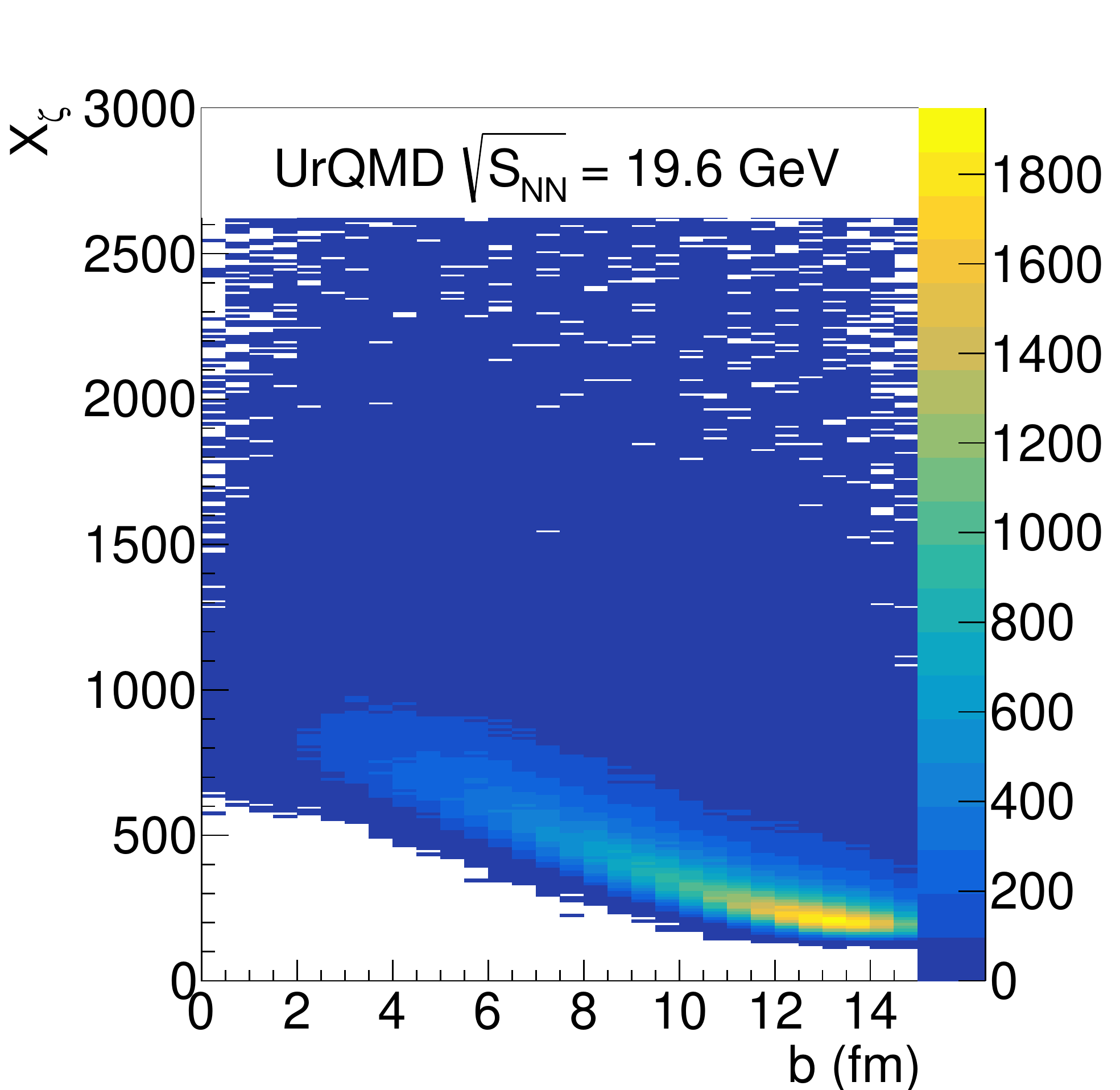}
    \caption{$X_{\zeta}$ value versus impact parameter for $\sqrt{S_{NN}}$ = 19.6 GeV UrQMD simulations. The extended tail of the distribution is due to Landau fluctuations.}
    \label{fig:NmipRaw}
\end{figure}

Figure~\ref{fig:NmipRaw} shows the (anti-)correlation between the impact parameter from the model and the sum of the signals ($X_{\zeta} \equiv \sum_i\zeta_i$) from the 744 EPD tiles for 19.6~GeV Au+Au collisions. The ``noise'' effect of Landau fluctuations is clear in the extended tail of the distribution at large $X_{\zeta}$, reducing the correlation.

\begin{figure}
    \centering
    \includegraphics[width=0.45\textwidth]{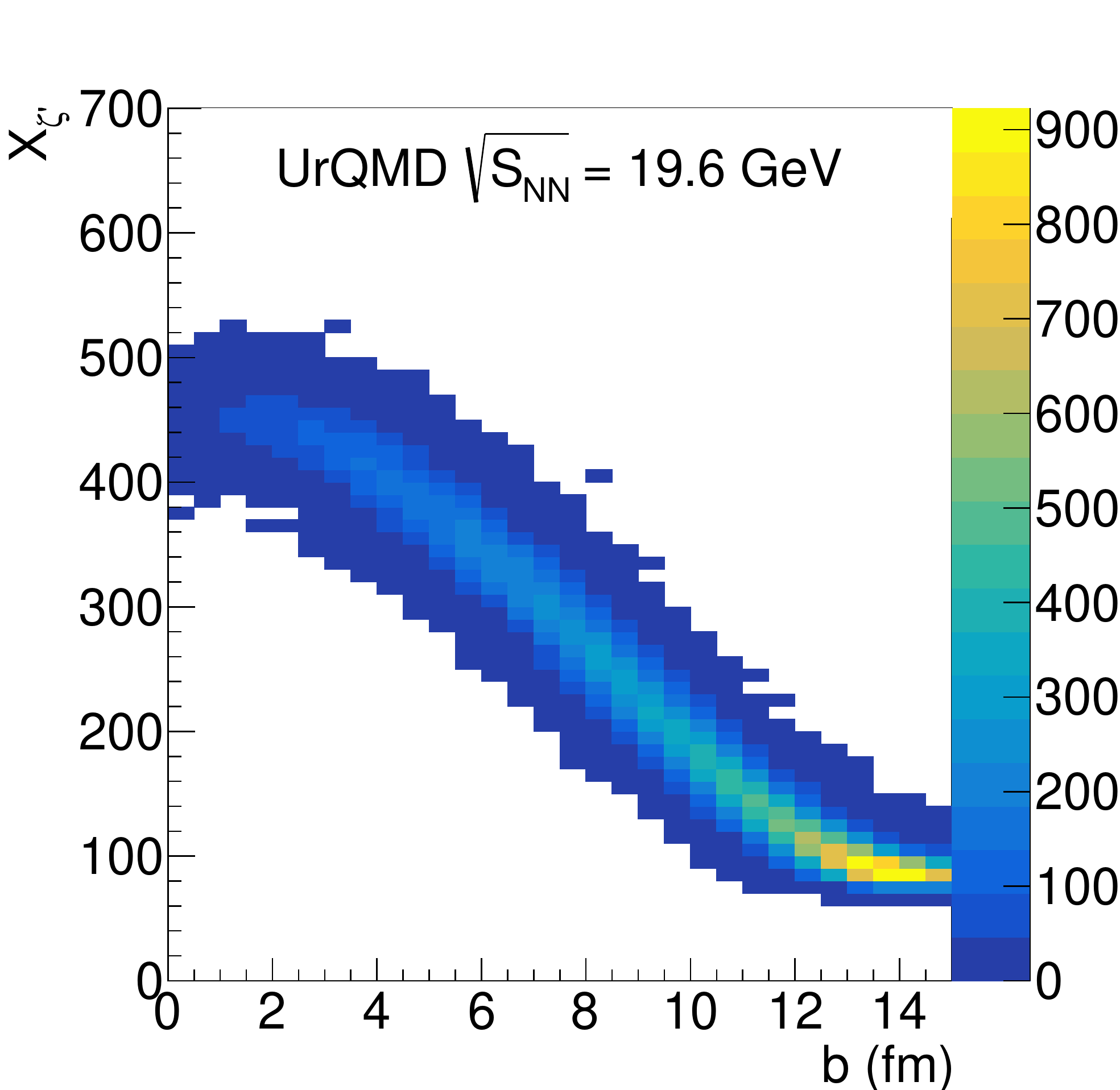}
    \caption{$X_{\zeta'}$ value versus impact parameter for $\sqrt{s_{NN}}$ = 19.6 GeV UrQMD simulations. The truncation of $\zeta$ (from Equation \ref{eq:zprime}) reduces the noise from Landau fluctuations seen in the correlation between $X_{\zeta}$ and $b$ in Figure \ref{fig:NmipRaw}.}
    \label{fig:TnMip}
\end{figure}

The effect of these fluctuations may be reduced by ``truncating'' the signal from each tile, replacing a tile's signal with:
\begin{equation}
      \zeta^\prime \equiv \begin{cases}
        \zeta, & \text{if~}\zeta < {\rm Mx} \\
    \textrm{Mx},              & \text{otherwise}
    \end{cases}
    \label{eq:zprime}
\end{equation}
We chose the value of ${\rm Mx} = 3$ for all energies and centralities for this paper, though this could be tuned based on an analysis of the most probable value of the number of particles that will pass through a given tile.  The result of applying this methodology can be seen in Figure~\ref{fig:TnMip}, which plots $X_{\zeta'} \equiv \sum_i\zeta^\prime_i$ versus impact parameter. The Pearson coefficient is about 0.98, as compared to only 0.39 for the correlation in Figure~\ref{fig:NmipRaw}.

\subsection{A new centrality estimator using ring weights}
\label{sec:RingWeights}
As we discuss in detail in section~\ref{sec:results}, the 16 rings of the EPD (corresponding to different $|\eta|$ ranges listed in Table \ref{table:rings}) can be affected quite differently as the impact parameter of the collision is varied. Indeed, depending on the collision energy, the signal in some rings may {\it increase} as $b$ is increased, while the signal in others may {\it decrease}. Thus their contributions to the simple sum $X_{\zeta'}$ discussed may partly cancel, reducing the sensitivity of this measure to collision centrality.

The differential response can be accounted for-- indeed, even be exploited to increase sensitivity-- by constructing a new simple measure that weights each ring's ``contribution'' differently.
Below, we consider two ways to quantify this contribution.
\subsubsection{Weighted sum of $\zeta'$ \label{sec:WeightsZeta}}
First, we define a ring's contribution to be the sum of the truncated signals in each tile in the ring:
\begin{equation}
\label{eq:RingContributionZetaPrime}
    C_r \equiv \sum_{\substack{\text{tile }j\text{ in}\\ \text{ring } r}}\zeta^\prime_j\qquad\text{(energy-loss based)}
\end{equation}
where we have indicated explicitly that the ring's contribution is based on energy loss (in the next section, we will discuss an analogous, particle based approach). Since we consider symmetric collisions occurring mid-way between the EPD wheels, ring $r$ on the wheel positioned at $z=-375$~cm is summed with ring $r$ on the wheel at $z=+375$~cm.

Based on these contributions, we define our centrality measure as a weighted sum:
\begin{equation}
\label{eq:Xzeta}
    X_{W,\zeta^\prime} \equiv \sum^{16}_{r=1} W_r C_r + W_{17}
\end{equation}
where $W_i$ are parameters determined below.

We wish to maximize the correlation between $X_{\rm \zeta'}$ and some global quantity $G$ (for the moment, $G$ is the impact parameter, $b$, but we generalize the discussion in section~\ref{sec:summary}). A figure of merit may be the squared residual, summed over events:
\begin{equation}\label{eqn:chi2}
    \chi^2 = \sum^{N_{events}}_{j=1} (X_{W,\zeta^\prime,j} - G_j)^2  ,
\end{equation}
where $X_{W,\zeta^\prime,j}$ and $G_j$ are respectively the values of the estimator and 
  global quantity (e.g. impact parameter) for event $j$.

Maximizing $\chi^2$ yields 17 linear equations:
\begin{equation}
\sum_{q=1}^{17} A_{q,t} W_q = B_t 
\end{equation}
where:
\begin{align}
    A_{q,t}    &= \sum_{j=1}^{N_{\rm events}} C_{q,j}C_{t,j} & {\rm for~}q,t=1...16 &~\\
    A_{17,t}   &= \sum_{j=1}^{N_{\rm events}}C_{t,j}         & {\rm for~}t=1...16 &~\\
    A_{17,17}  &= N_{\rm events} &~&~\\
    B_{t}    &= \sum_{j=1}^{N_{events}}G_j C_{t,j} & {\rm for~}t=1...16 &~ \label{eq:Bt}\\
    B_{17}   &= \sum_{j=1}^{N_{events}}G_j &~ &~ \label{eq:B17}
\end{align}
Since $A$ is a symmetric, real, $17\times17$ matrix it can be easily inverted to find the unique best parameters $W_t$.
\subsubsection{Weighted sum of particles}
As discussed above, an EPD tile measures in a given event the energy deposited in the tile, and not the number of particles that actually passed through the tile. However, it is natural to ask whether the sensitivity would be improved if a ring's contribution would be the number of charged particles passing through tiles in that ring, i.e. if $C_r$ in equation~\ref{eq:RingContributionZetaPrime} would be redefined: 
\begin{equation}
\label{eq:RingContributionParticle}
    C_r \equiv \sum_{\substack{\text{tile }j\text{ in}\\ \text{ring } r}}N_j\qquad\text{(particle based)}
\end{equation}
Here, $N_j$ is the number of particles that passed through tile $j$ in the event.

Analogous to $X_{W,\zeta^\prime}$ a weighted-sum centrality measure $X_{W,FWD}$ may then be constructed from the contributions of equation~\ref{eq:RingContributionParticle}, with the weights determined by equations~\ref{eqn:chi2}-\ref{eq:B17}.
 
\section{\label{sec:results}Results}

\begin{figure}
    \centering
    \includegraphics[width=0.45\textwidth]{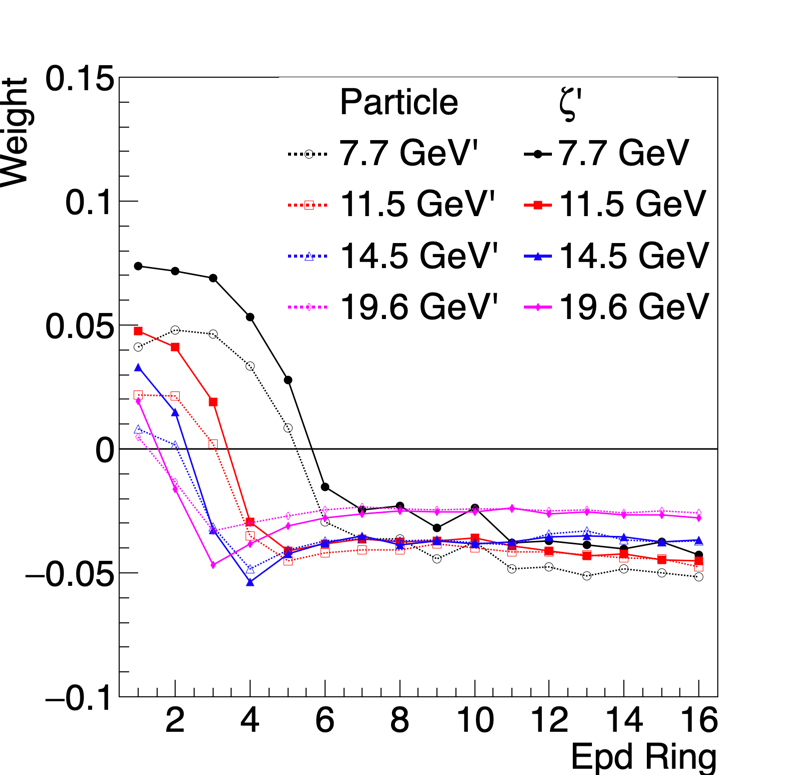}
    \caption{The weights for both forward $\eta$ particles and $\zeta'$ when using the linear weight scheme from Section \ref{sec:RingWeights}, by EPD ring. The sign change in ring weights is motivated by spectator proton intrusion into the EPD's acceptance window, as can be seen in Figure \ref{fig:Rapidity}.}
    \label{fig:Weights}
\end{figure}

With the mathematical formalism described in Section \ref{sec:RingWeights}, we can examine the performance of the EPD as it pertains to relating signals within the detector to the impact parameter using UrQMD.  

\subsection{\label{sec:level2}Correlations between yields and $b$}

\begin{figure}
    \centering
    \includegraphics[width=0.45\textwidth]{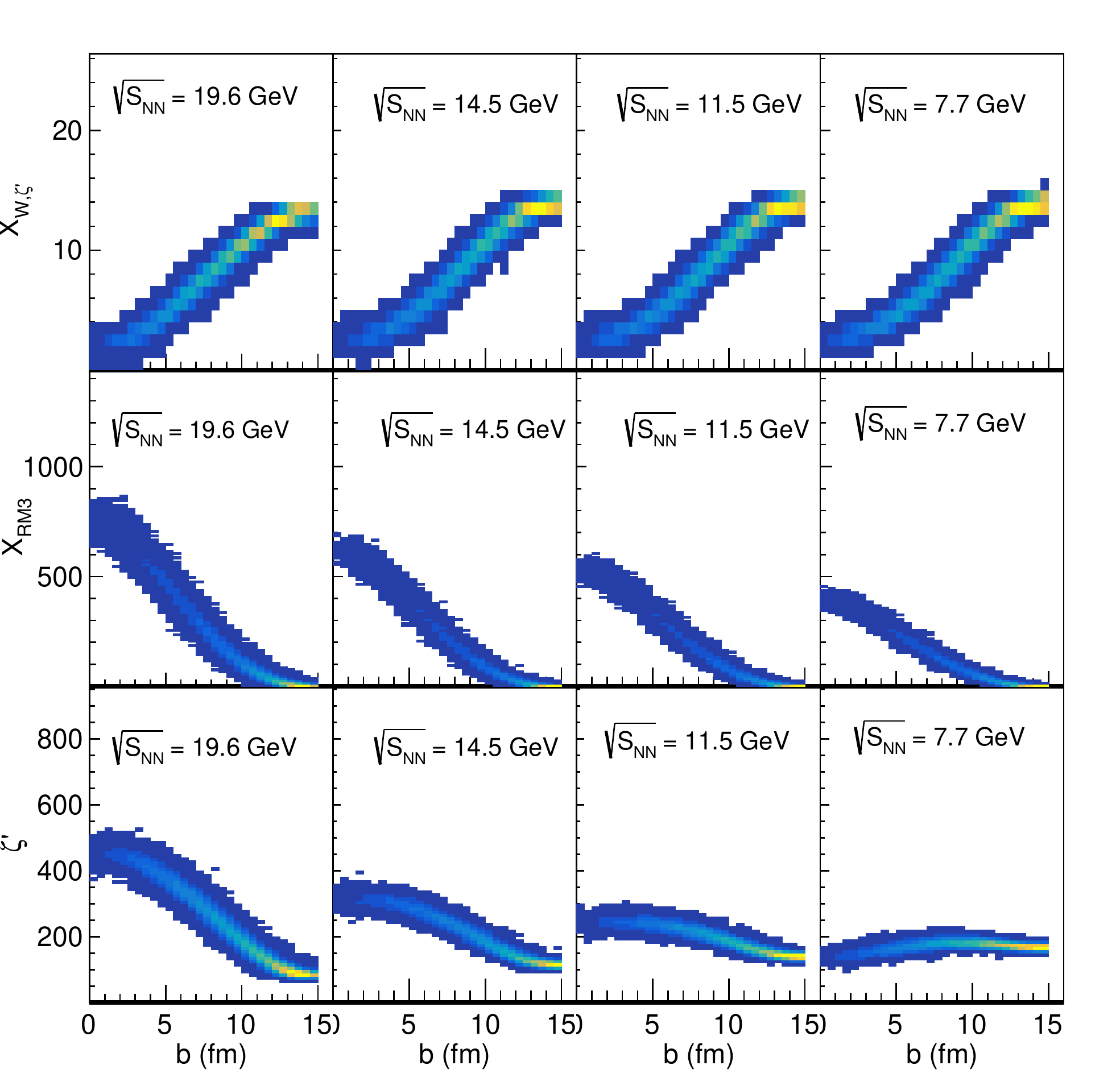}
    \caption{Correlation between the impact parameter, $b$, and three observables: $X_{W,\zeta'}$ (top), $X_{RM3}$ (middle), $X_{\zeta'}$ (bottom), for four different collision energies. 
        } 
    \label{fig:Variations}
\end{figure}

In order to apply the formalism described in the previous section, we will compare the two mid-rapidity observables ($X_{RM1}$ and $X_{RM3}$) with four forward rapidity observables ($X_{\zeta^\prime}, X_{W,\zeta^\prime}, X_{FWD}$, and $X_{W,FWD}$), where only the observables designated by $W$ use the linear weight method formalized above. The weights determined by this method versus EPD ring number can be seen in Figure \ref{fig:Weights}, with the acceptance of the EPD rings detailed in Table \ref{table:rings}.  The sign of the weights changes when moving from a distribution that is dominated by spectators versus participants.  In Figure \ref{fig:Weights}, we can see that the EPD ring this occurs in changes as the collision energy changes due to the change in beam rapidity (see Figure \ref{fig:Rapidity}).  The weights for EPD distributions with $\zeta^\prime$ are very similar to those required by the raw particle counts.  These weights were then used to determine the correlation between $\{X_{W,FWD}, X_{W,\zeta^\prime}\}$ and $b$, the results of which can be seen in Figure \ref{fig:Variations}.

Figure \ref{fig:FwdAllParticle} is in agreement with the conclusions of the Chatterjee analysis \cite{Chatterjee:2019fey}, and indicates that a summation of the yield over the EPD acceptance is a poor observable. However, when we apply the linear weighting technique described in Section \ref{sec:RingWeights}, we recover a much more usable correlation between forward $\eta$ particle yields and $b$. This is summarised in Figure \ref{fig:Variations}.  For all the energies under consideration, $X_{RM3}$ (middle row) is well correlated with the impact parameter, whereas  $X_{\zeta'}$ from the EPD (bottom row) shows a decreasing correlation with decreasing energy.  However, the linear weighted $X_{W,\zeta'}$ (top row) shows that the correlation is restored.

Experimentally, the centrality is not determined by a model dependent relationship between a global observable and the impact parameter, but rather by considering the global observable's distribution quantiles (though it should be noted that real world analyses include a Glauber model as the efficiency of recording an event only approaches 100\% for the most central collisions~\cite{Miller:2007ri}).  This indicates that, as long as a global observable is reasonably correlated with the impact parameter, the centrality distribution based on this selection criteria will also be reasonable.

\begin{figure}
    \centering
    \includegraphics[width=0.45\textwidth]{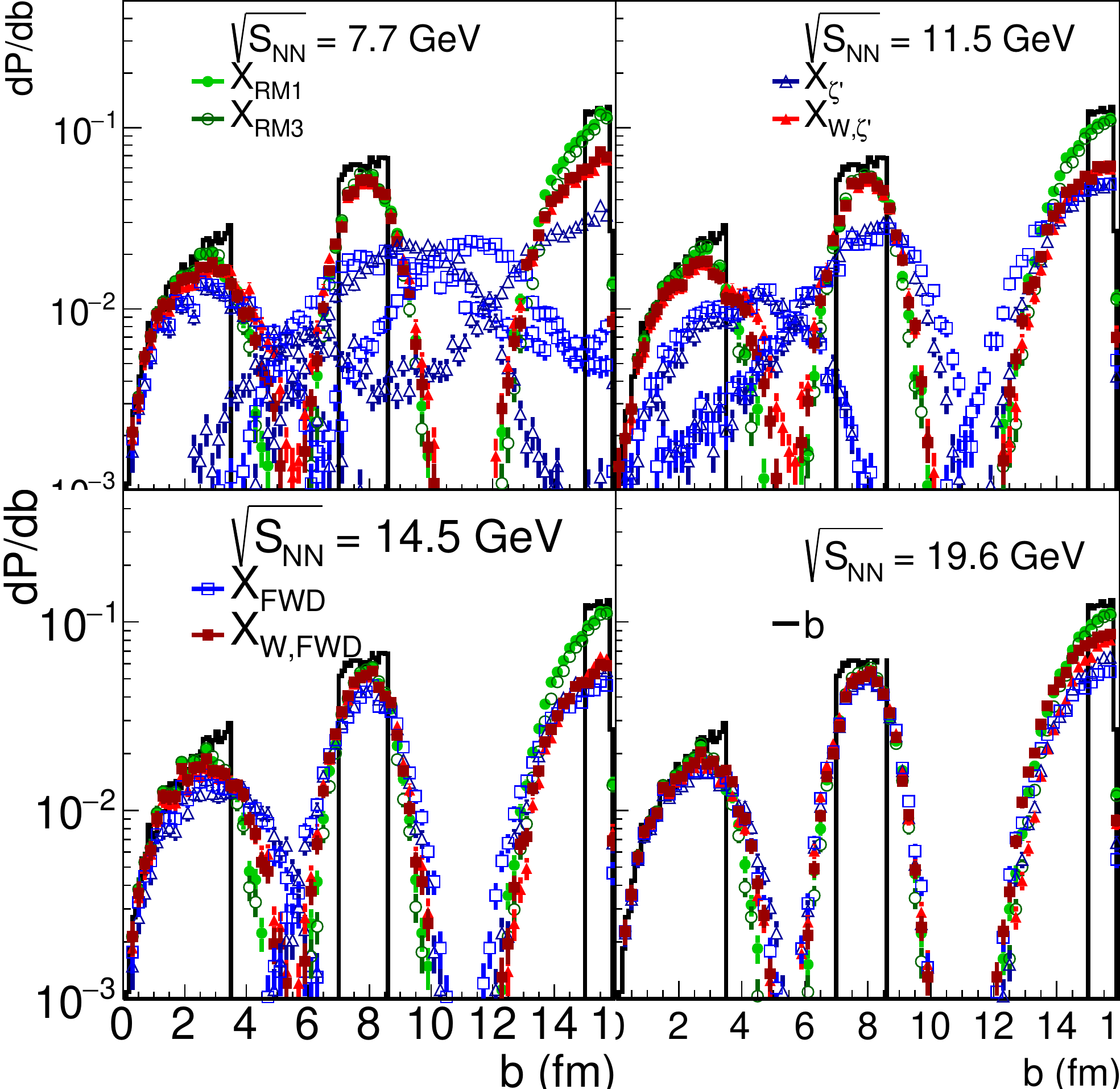}
    \caption{The impact parameter distributions for centrality selections 0 - 5\%, 20 - 30\% and 90 - 100\%.  The black histograms are the $b$ impact parameter distributions if the centrality selection is determined from the impact parameter directly.  The green circles are determined using $X_{RM1}$ (closed) and $X_{RM3}$ (open), the triangles are determined using $X_{\zeta'}$ (open blue) and $X_{W,\zeta'}$ (closed red), and the squares are determined using $X_{FWD}$ (open blue) and $X_{W,FWD}$ (closed red).
    } 
    \label{fig:ImpSelection}
\end{figure}

In Figure \ref{fig:ImpSelection}, the impact parameter distributions that are determined by using the appropriate quantiles for the global variables are shown along with distributions that result from directly using the quantiles of the impact parameter distribution.  For $\sqrt{s_{NN}}$= 19.6 GeV, all methods performed similarly (quantified in Section \ref{sec:resRes}).  The mid-rapidity observables, $X_{RM1}$ and $X_{RM3}$, have $b$ distributions which peak within the $b$ distribution slices and do not change drastically as the collision energy decreases from 19.6 GeV to 7.7 GeV.  The forward observables without any weighting, $X_{FWD}$ and $X_{\zeta'}$, have distributions which no longer lie under the $b$ distribution slices at the lower energies; this suggests a poor centrality resolution for these observables in the lower energy ranges of the BES program. This potential loss in resolution, however, can be compensated for by applying the weighting scheme discussed in section \ref{sec:methods}. We see that the distributions for $X_{W,FWD}$ and $X_{W,\zeta'}$ are under the $b$ distribution slices for all collision energies under consideration.

\begin{figure}
    \centering
    \includegraphics[width=0.45\textwidth]{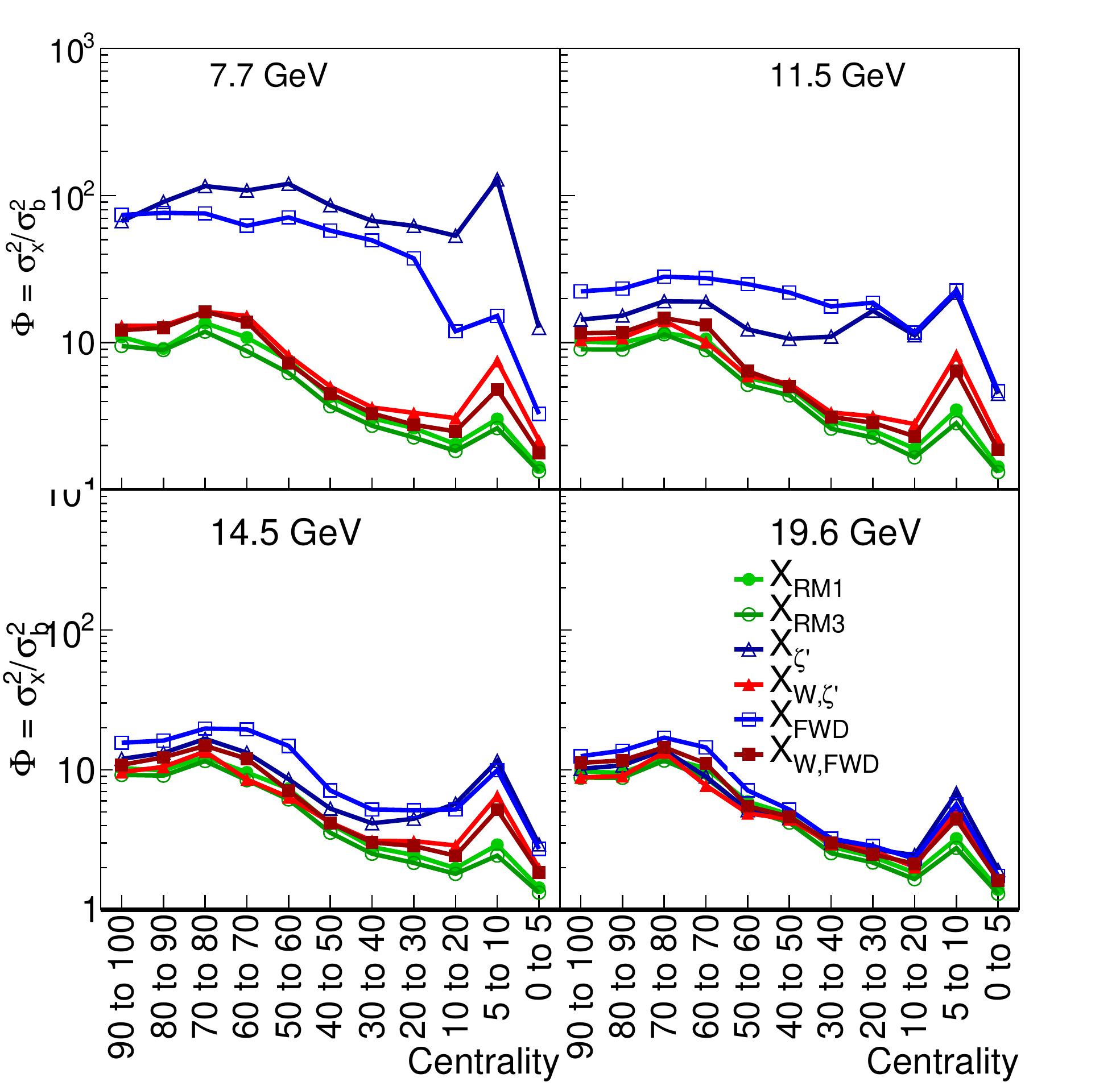}
    \caption{Centrality resolution for collision energies $\sqrt{s_{NN}}$ = 19.6, 14.5, 11.5, and 7.7 GeV.  The observables based on mid-rapidity multiplicity are $X_{RM1}$ (green closed circles) and $X_{RM3}$ (dark green open circles).  The observables based on unweighted distributions from the forward region are $X_{FWD}$ (blue open square) for particle yield and $X_{\zeta'}$ (blue open triangle) for truncated energy loss in the EPD.  The observables based on linear weights are $X_{W,FWD}$ (closed dark red square) and $X_{W,\zeta'}$ (closed red triangle).
    }
    \label{fig:CentralityResolution}
\end{figure}

\subsection{\label{sec:resRes}Centrality Resolution}
From the $b$ distributions in Figure \ref{fig:ImpSelection}, we determined the centrality resolution for all $X$. As in \cite{Chatterjee:2019fey}, we employ the centrality resolution metric $\Phi$:

\begin{equation}
    \Phi_i \equiv \frac{\sigma_{b_{X,i}}^2}{\sigma_{b_i}^2}
\end{equation}
where $\sigma_{b_i}^2$ is the variance from the impact parameter distribution for a given centrality range $i$ when centrality is determined by $b$, and $\sigma_{b_{X,i}}^2$ is the variance from the impact parameter distribution for the same centrality range $i$ where centrality is determined by $X$.

Results for $\Phi$ can be seen in Figure \ref{fig:CentralityResolution}.  At the top RHIC energies, all centrality methods perform similarly.  The resolution at lowest energies is poorest for unweighted distributions in the forward region ($X_{FWD}$ for particle yield, and $X_{\zeta'}$ for truncated energy loss in the EPD), but this resolution is clearly recovered when we apply the linear weights detailed in Section \ref{sec:RingWeights} ($X_{W,FWD}$ and $X_{W,\zeta'}$, respectively). Both $X_{RM1}$ and $X_{RM3}$ are based on mid-rapidity multiplicities. The anomalous, upward point at the 5-10\% range is due to the smaller centrality bins used for our two most central selections, which increases the standard deviation of the distributions.

The poor performance of $X_{FWD}$ and $X_{\zeta'}$, which agrees with the conclusions in \cite{Chatterjee:2019fey}, is due to spectator proton intrusion into the EPD's acceptance. If we do not weight the EPD rings which are dominated by the spectator protons yield, which have positive correlation with $b$, with a different sign than the EPD rings where we are dominate by participants, which have a negative correlation, they will cancel out. This is the entire purpose of the methodology in section \ref{sec:RingWeights}; thus the correlation weights found for rings with spectators will have inverted signs compared with those rings with only participants (Figures \ref{fig:Rapidity} and \ref{fig:Weights}). The linear weight method recovers the centrality resolution lost by the simple sum method of $X_{FWD}$ and $X_{\zeta'}$

\section{\label{sec:summary}Summary and Discussion}
Observables in heavy-ion experiments can suffer from auto-correlation effects if the particles used to construct the observable are from the same acceptance as the event centrality is defined (e.g. $\kappa$ and $S$ for net-proton multiplicity being studied as part of the STAR BES program)~\cite{Luo:2013bmi}. For observables analysed at mid-rapidity, such as those measured using the TPC in the STAR experiment, it would thus be preferable to select centrality using a forward detector. The EPD is the prime candidate for a forward centrality selection detector at STAR, but at the lower collision energies of the BES there is significant spectator proton intrusion into the $\eta$ acceptance window of the EPD. It had been suggested that this spectator intrusion could potentially degrade the centrality resolution of the EPD due to the inverse correlation effect on particle yields with $b$ in those $\eta$ regions where the EPD and spectator protons coincide \cite{Chatterjee:2019fey}.  However we showed that by treating the particle yield correlations from the EPD rings individually, instead of simply summing the yield from the total EPD acceptance, we can account for correlations with both participants and spectators in an event. This treatment leverages the spectator protons in the EPD acceptance range as a relevant marker for global quantities (such as $b$).

In this paper, we outlined a method of applying a linear weight to the rings (Equation \ref{eq:Xzeta}) as one example of a procedure that improves the centrality resolution by properly weighting the contribution from each EPD ring based on minimizing the residual between a global observable and the EPD rings. Results from this method were shown for correlations with $b$, but the method is sufficiently general that weighting may be found for any global quantity $G$; for instance, the method could also be employed to weigh the EPD ring contributions as they correlate with $X_{RM3}$ or V. 

More sophisticated methods than the simplest one we described in section~\ref{sec:RingWeights} are certainly
  possible and are under development.

Using this method we recovered the centrality resolution in the forward $\eta$ region that is lost in the lower energy ranges of STAR BES when simply considering a sum of the yields over the entire EPD $\eta$ range. Further, EPD simulation shows no marked degradation of centrality resolution when comparing centrality using the linear weighted sum of particles in the EPD acceptance range ($X_{W,Fwd}$) versus simulated energy deposition in the EPD itself ({$X_{W,\zeta'}$}). This strongly suggests the EPD can be used as a reliable centrality detector in STAR BES energy ranges of $\sqrt{s_{NN}}$ = 7.7, 11.5, 14.5, 19.6, and 200 GeV, which would greatly reduce the possibility of autocorrelations in analyses of observables at mid-rapidity.

We conclude by briefly considering the implication of this study in an experimental analysis,
  in which the true impact parameter is unknown.
Any estimate of the impact parameter resolution of any measurable estimator is then completely model-dependent;
  $X_{\zeta^\prime}$ may in fact be the best estimator in reality, despite being the worst in the
  UrQMD calculations (c.f. figure~\ref{fig:CentralityResolution}).
In such a case-- especially if even a years-long program of high-quality data results in a subtle wiggle in fluctuations
  at only the $3\sigma$ level~\cite{Adam:2020unf}-- nontrivial effects of autocorrelations must be ruled
  out in a model-independent way, not relying solely on transport calculations~\cite{Chatterjee:2019fey}.
The experimental procedure must be to 
  (1) perform the analysis, using a common estimator (e.g. $X_{RM3}$) to select on centrality;
  (2) in a kinematic region far from the fluctuation measurement, construct a weighted estimator by following the procedure of 
      section~\ref{sec:WeightsZeta}, where the common estimator is used as the global quantity $G$ in equations~\ref{eq:Bt}-\ref{eq:B17};
  (3) repeat the analysis, using the new estimator (e.g. $X_{W,\zeta^\prime}$) to select on centrality.
Persistence of the signal when using the new estimator would lead to greater confidence that  autocorrelations
  are not influencing the signal itself.
Such confidence would be most welcome for the subtlest signals of fundamental physics of QCD.

\section*{Acknowledgements}
We thank Xiaofeng Luo and Arghya Chatterjee for fruitful discussions.
This work supported by the U.S. National Science Foundation grant 1945296 and the U.S. Department of Energy grant DE-SC0020651.

\bibliography{EPDcentrality}

\begin{thebibliography}{34}%
\makeatletter
\providecommand \@ifxundefined [1]{%
 \@ifx{#1\undefined}
}%
\providecommand \@ifnum [1]{%
 \ifnum #1\expandafter \@firstoftwo
 \else \expandafter \@secondoftwo
 \fi
}%
\providecommand \@ifx [1]{%
 \ifx #1\expandafter \@firstoftwo
 \else \expandafter \@secondoftwo
 \fi
}%
\providecommand \natexlab [1]{#1}%
\providecommand \enquote  [1]{``#1''}%
\providecommand \bibnamefont  [1]{#1}%
\providecommand \bibfnamefont [1]{#1}%
\providecommand \citenamefont [1]{#1}%
\providecommand \href@noop [0]{\@secondoftwo}%
\providecommand \href [0]{\begingroup \@sanitize@url \@href}%
\providecommand \@href[1]{\@@startlink{#1}\@@href}%
\providecommand \@@href[1]{\endgroup#1\@@endlink}%
\providecommand \@sanitize@url [0]{\catcode `\\12\catcode `\$12\catcode
  `\&12\catcode `\#12\catcode `\^12\catcode `\_12\catcode `\%12\relax}%
\providecommand \@@startlink[1]{}%
\providecommand \@@endlink[0]{}%
\providecommand \url  [0]{\begingroup\@sanitize@url \@url }%
\providecommand \@url [1]{\endgroup\@href {#1}{\urlprefix }}%
\providecommand \urlprefix  [0]{URL }%
\providecommand \Eprint [0]{\href }%
\providecommand \doibase [0]{http://dx.doi.org/}%
\providecommand \selectlanguage [0]{\@gobble}%
\providecommand \bibinfo  [0]{\@secondoftwo}%
\providecommand \bibfield  [0]{\@secondoftwo}%
\providecommand \translation [1]{[#1]}%
\providecommand \BibitemOpen [0]{}%
\providecommand \bibitemStop [0]{}%
\providecommand \bibitemNoStop [0]{.\EOS\space}%
\providecommand \EOS [0]{\spacefactor3000\relax}%
\providecommand \BibitemShut  [1]{\csname bibitem#1\endcsname}%
\let\auto@bib@innerbib\@empty
\bibitem [{\citenamefont {Chatterjee}\ \emph {et~al.}(2020)\citenamefont
  {Chatterjee}, \citenamefont {Zhang}, \citenamefont {Zeng}, \citenamefont
  {Sahoo},\ and\ \citenamefont {Luo}}]{Chatterjee:2019fey}%
  \BibitemOpen
  \bibfield  {author} {\bibinfo {author} {\bibfnamefont {A.}~\bibnamefont
  {Chatterjee}}, \bibinfo {author} {\bibfnamefont {Y.}~\bibnamefont {Zhang}},
  \bibinfo {author} {\bibfnamefont {J.}~\bibnamefont {Zeng}}, \bibinfo {author}
  {\bibfnamefont {N.~R.}\ \bibnamefont {Sahoo}}, \ and\ \bibinfo {author}
  {\bibfnamefont {X.}~\bibnamefont {Luo}},\ }\href {\doibase
  10.1103/PhysRevC.101.034902} {\bibfield  {journal} {\bibinfo  {journal}
  {Phys. Rev. C}\ }\textbf {\bibinfo {volume} {101}},\ \bibinfo {pages}
  {034902} (\bibinfo {year} {2020})},\ \Eprint
  {http://arxiv.org/abs/1910.08004} {arXiv:1910.08004 [nucl-ex]} \BibitemShut
  {NoStop}%
\bibitem [{\citenamefont {Adams}\ \emph {et~al.}(2005)\citenamefont {Adams}
  \emph {et~al.}}]{Adams:2005dq}%
  \BibitemOpen
  \bibfield  {author} {\bibinfo {author} {\bibfnamefont {J.}~\bibnamefont
  {Adams}} \emph {et~al.} (\bibinfo {collaboration} {STAR}),\ }\href {\doibase
  10.1016/j.nuclphysa.2005.03.085} {\bibfield  {journal} {\bibinfo  {journal}
  {Nucl. Phys. A}\ }\textbf {\bibinfo {volume} {757}},\ \bibinfo {pages} {102}
  (\bibinfo {year} {2005})},\ \Eprint {http://arxiv.org/abs/nucl-ex/0501009}
  {arXiv:nucl-ex/0501009} \BibitemShut {NoStop}%
\bibitem [{\citenamefont {Arsene}\ \emph {et~al.}(2005)\citenamefont {Arsene}
  \emph {et~al.}}]{Arsene:2004fa}%
  \BibitemOpen
  \bibfield  {author} {\bibinfo {author} {\bibfnamefont {I.}~\bibnamefont
  {Arsene}} \emph {et~al.} (\bibinfo {collaboration} {BRAHMS}),\ }\href
  {\doibase 10.1016/j.nuclphysa.2005.02.130} {\bibfield  {journal} {\bibinfo
  {journal} {Nucl. Phys. A}\ }\textbf {\bibinfo {volume} {757}},\ \bibinfo
  {pages} {1} (\bibinfo {year} {2005})},\ \Eprint
  {http://arxiv.org/abs/nucl-ex/0410020} {arXiv:nucl-ex/0410020} \BibitemShut
  {NoStop}%
\bibitem [{\citenamefont {Back}\ \emph {et~al.}(2005)\citenamefont {Back} \emph
  {et~al.}}]{Back:2004je}%
  \BibitemOpen
  \bibfield  {author} {\bibinfo {author} {\bibfnamefont {B.}~\bibnamefont
  {Back}} \emph {et~al.} (\bibinfo {collaboration} {PHOBOS}),\ }\href {\doibase
  10.1016/j.nuclphysa.2005.03.084} {\bibfield  {journal} {\bibinfo  {journal}
  {Nucl. Phys. A}\ }\textbf {\bibinfo {volume} {757}},\ \bibinfo {pages} {28}
  (\bibinfo {year} {2005})},\ \Eprint {http://arxiv.org/abs/nucl-ex/0410022}
  {arXiv:nucl-ex/0410022} \BibitemShut {NoStop}%
\bibitem [{\citenamefont {Adcox}\ \emph {et~al.}(2005)\citenamefont {Adcox}
  \emph {et~al.}}]{Adcox:2004mh}%
  \BibitemOpen
  \bibfield  {author} {\bibinfo {author} {\bibfnamefont {K.}~\bibnamefont
  {Adcox}} \emph {et~al.} (\bibinfo {collaboration} {PHENIX}),\ }\href
  {\doibase 10.1016/j.nuclphysa.2005.03.086} {\bibfield  {journal} {\bibinfo
  {journal} {Nucl. Phys. A}\ }\textbf {\bibinfo {volume} {757}},\ \bibinfo
  {pages} {184} (\bibinfo {year} {2005})},\ \Eprint
  {http://arxiv.org/abs/nucl-ex/0410003} {arXiv:nucl-ex/0410003} \BibitemShut
  {NoStop}%
\bibitem [{\citenamefont {Stephanov}(2006)}]{Stephanov:2007fk}%
  \BibitemOpen
  \bibfield  {author} {\bibinfo {author} {\bibfnamefont {M.}~\bibnamefont
  {Stephanov}},\ }\href {\doibase 10.22323/1.032.0024} {\bibfield  {journal}
  {\bibinfo  {journal} {PoS}\ }\textbf {\bibinfo {volume} {LAT2006}},\ \bibinfo
  {pages} {024} (\bibinfo {year} {2006})},\ \Eprint
  {http://arxiv.org/abs/hep-lat/0701002} {arXiv:hep-lat/0701002} \BibitemShut
  {NoStop}%
\bibitem [{\citenamefont {Bowman}\ and\ \citenamefont
  {Kapusta}(2009)}]{Bowman:2008kc}%
  \BibitemOpen
  \bibfield  {author} {\bibinfo {author} {\bibfnamefont {E.}~\bibnamefont
  {Bowman}}\ and\ \bibinfo {author} {\bibfnamefont {J.~I.}\ \bibnamefont
  {Kapusta}},\ }\href {\doibase 10.1103/PhysRevC.79.015202} {\bibfield
  {journal} {\bibinfo  {journal} {Phys. Rev. C}\ }\textbf {\bibinfo {volume}
  {79}},\ \bibinfo {pages} {015202} (\bibinfo {year} {2009})},\ \Eprint
  {http://arxiv.org/abs/0810.0042} {arXiv:0810.0042 [nucl-th]} \BibitemShut
  {NoStop}%
\bibitem [{\citenamefont {Aoki}\ \emph {et~al.}(2006)\citenamefont {Aoki},
  \citenamefont {Endrodi}, \citenamefont {Fodor}, \citenamefont {Katz},\ and\
  \citenamefont {Szabo}}]{Aoki:2006we}%
  \BibitemOpen
  \bibfield  {author} {\bibinfo {author} {\bibfnamefont {Y.}~\bibnamefont
  {Aoki}}, \bibinfo {author} {\bibfnamefont {G.}~\bibnamefont {Endrodi}},
  \bibinfo {author} {\bibfnamefont {Z.}~\bibnamefont {Fodor}}, \bibinfo
  {author} {\bibfnamefont {S.}~\bibnamefont {Katz}}, \ and\ \bibinfo {author}
  {\bibfnamefont {K.}~\bibnamefont {Szabo}},\ }\href {\doibase
  10.1038/nature05120} {\bibfield  {journal} {\bibinfo  {journal} {Nature}\
  }\textbf {\bibinfo {volume} {443}},\ \bibinfo {pages} {675} (\bibinfo {year}
  {2006})},\ \Eprint {http://arxiv.org/abs/hep-lat/0611014}
  {arXiv:hep-lat/0611014} \BibitemShut {NoStop}%
\bibitem [{\citenamefont {Gupta}\ \emph {et~al.}(2011)\citenamefont {Gupta},
  \citenamefont {Luo}, \citenamefont {Mohanty}, \citenamefont {Ritter},\ and\
  \citenamefont {Xu}}]{Gupta:2011wh}%
  \BibitemOpen
  \bibfield  {author} {\bibinfo {author} {\bibfnamefont {S.}~\bibnamefont
  {Gupta}}, \bibinfo {author} {\bibfnamefont {X.}~\bibnamefont {Luo}}, \bibinfo
  {author} {\bibfnamefont {B.}~\bibnamefont {Mohanty}}, \bibinfo {author}
  {\bibfnamefont {H.~G.}\ \bibnamefont {Ritter}}, \ and\ \bibinfo {author}
  {\bibfnamefont {N.}~\bibnamefont {Xu}},\ }\href {\doibase
  10.1126/science.1204621} {\bibfield  {journal} {\bibinfo  {journal}
  {Science}\ }\textbf {\bibinfo {volume} {332}},\ \bibinfo {pages} {1525}
  (\bibinfo {year} {2011})},\ \Eprint {http://arxiv.org/abs/1105.3934}
  {arXiv:1105.3934 [hep-ph]} \BibitemShut {NoStop}%
\bibitem [{\citenamefont {Fodor}\ and\ \citenamefont
  {Katz}(2004)}]{Fodor:2004nz}%
  \BibitemOpen
  \bibfield  {author} {\bibinfo {author} {\bibfnamefont {Z.}~\bibnamefont
  {Fodor}}\ and\ \bibinfo {author} {\bibfnamefont {S.}~\bibnamefont {Katz}},\
  }\href {\doibase 10.1088/1126-6708/2004/04/050} {\bibfield  {journal}
  {\bibinfo  {journal} {JHEP}\ }\textbf {\bibinfo {volume} {04}},\ \bibinfo
  {pages} {050} (\bibinfo {year} {2004})},\ \Eprint
  {http://arxiv.org/abs/hep-lat/0402006} {arXiv:hep-lat/0402006} \BibitemShut
  {NoStop}%
\bibitem [{\citenamefont {de~Forcrand}\ and\ \citenamefont
  {Philipsen}(2002)}]{deForcrand:2002hgr}%
  \BibitemOpen
  \bibfield  {author} {\bibinfo {author} {\bibfnamefont {P.}~\bibnamefont
  {de~Forcrand}}\ and\ \bibinfo {author} {\bibfnamefont {O.}~\bibnamefont
  {Philipsen}},\ }\href {\doibase 10.1016/S0550-3213(02)00626-0} {\bibfield
  {journal} {\bibinfo  {journal} {Nucl. Phys. B}\ }\textbf {\bibinfo {volume}
  {642}},\ \bibinfo {pages} {290} (\bibinfo {year} {2002})},\ \Eprint
  {http://arxiv.org/abs/hep-lat/0205016} {arXiv:hep-lat/0205016} \BibitemShut
  {NoStop}%
\bibitem [{\citenamefont {Qin}\ \emph {et~al.}(2011)\citenamefont {Qin},
  \citenamefont {Chang}, \citenamefont {Chen}, \citenamefont {Liu},\ and\
  \citenamefont {Roberts}}]{Qin:2010nq}%
  \BibitemOpen
  \bibfield  {author} {\bibinfo {author} {\bibfnamefont {S.-x.}\ \bibnamefont
  {Qin}}, \bibinfo {author} {\bibfnamefont {L.}~\bibnamefont {Chang}}, \bibinfo
  {author} {\bibfnamefont {H.}~\bibnamefont {Chen}}, \bibinfo {author}
  {\bibfnamefont {Y.-x.}\ \bibnamefont {Liu}}, \ and\ \bibinfo {author}
  {\bibfnamefont {C.~D.}\ \bibnamefont {Roberts}},\ }\href {\doibase
  10.1103/PhysRevLett.106.172301} {\bibfield  {journal} {\bibinfo  {journal}
  {Phys. Rev. Lett.}\ }\textbf {\bibinfo {volume} {106}},\ \bibinfo {pages}
  {172301} (\bibinfo {year} {2011})},\ \Eprint {http://arxiv.org/abs/1011.2876}
  {arXiv:1011.2876 [nucl-th]} \BibitemShut {NoStop}%
\bibitem [{\citenamefont {Xin}\ \emph {et~al.}(2014)\citenamefont {Xin},
  \citenamefont {Qin},\ and\ \citenamefont {Liu}}]{Xin:2014ela}%
  \BibitemOpen
  \bibfield  {author} {\bibinfo {author} {\bibfnamefont {X.-y.}\ \bibnamefont
  {Xin}}, \bibinfo {author} {\bibfnamefont {S.-x.}\ \bibnamefont {Qin}}, \ and\
  \bibinfo {author} {\bibfnamefont {Y.-x.}\ \bibnamefont {Liu}},\ }\href
  {\doibase 10.1103/PhysRevD.90.076006} {\bibfield  {journal} {\bibinfo
  {journal} {Phys. Rev. D}\ }\textbf {\bibinfo {volume} {90}},\ \bibinfo
  {pages} {076006} (\bibinfo {year} {2014})}\BibitemShut {NoStop}%
\bibitem [{\citenamefont {Shi}\ \emph {et~al.}(2014)\citenamefont {Shi},
  \citenamefont {Wang}, \citenamefont {Jiang}, \citenamefont {Cui},\ and\
  \citenamefont {Zong}}]{Shi:2014zpa}%
  \BibitemOpen
  \bibfield  {author} {\bibinfo {author} {\bibfnamefont {C.}~\bibnamefont
  {Shi}}, \bibinfo {author} {\bibfnamefont {Y.-L.}\ \bibnamefont {Wang}},
  \bibinfo {author} {\bibfnamefont {Y.}~\bibnamefont {Jiang}}, \bibinfo
  {author} {\bibfnamefont {Z.-F.}\ \bibnamefont {Cui}}, \ and\ \bibinfo
  {author} {\bibfnamefont {H.-S.}\ \bibnamefont {Zong}},\ }\href {\doibase
  10.1007/JHEP07(2014)014} {\bibfield  {journal} {\bibinfo  {journal} {JHEP}\
  }\textbf {\bibinfo {volume} {07}},\ \bibinfo {pages} {014} (\bibinfo {year}
  {2014})},\ \Eprint {http://arxiv.org/abs/1403.3797} {arXiv:1403.3797
  [hep-ph]} \BibitemShut {NoStop}%
\bibitem [{\citenamefont {Fischer}\ \emph {et~al.}(2014)\citenamefont
  {Fischer}, \citenamefont {Luecker},\ and\ \citenamefont
  {Welzbacher}}]{Fischer:2014ata}%
  \BibitemOpen
  \bibfield  {author} {\bibinfo {author} {\bibfnamefont {C.~S.}\ \bibnamefont
  {Fischer}}, \bibinfo {author} {\bibfnamefont {J.}~\bibnamefont {Luecker}}, \
  and\ \bibinfo {author} {\bibfnamefont {C.~A.}\ \bibnamefont {Welzbacher}},\
  }\href {\doibase 10.1103/PhysRevD.90.034022} {\bibfield  {journal} {\bibinfo
  {journal} {Phys. Rev. D}\ }\textbf {\bibinfo {volume} {90}},\ \bibinfo
  {pages} {034022} (\bibinfo {year} {2014})},\ \Eprint
  {http://arxiv.org/abs/1405.4762} {arXiv:1405.4762 [hep-ph]} \BibitemShut
  {NoStop}%
\bibitem [{\citenamefont {Lu}\ \emph {et~al.}(2015)\citenamefont {Lu},
  \citenamefont {Du}, \citenamefont {Cui},\ and\ \citenamefont
  {Zong}}]{Lu:2015naa}%
  \BibitemOpen
  \bibfield  {author} {\bibinfo {author} {\bibfnamefont {Y.}~\bibnamefont
  {Lu}}, \bibinfo {author} {\bibfnamefont {Y.-L.}\ \bibnamefont {Du}}, \bibinfo
  {author} {\bibfnamefont {Z.-F.}\ \bibnamefont {Cui}}, \ and\ \bibinfo
  {author} {\bibfnamefont {H.-S.}\ \bibnamefont {Zong}},\ }\href {\doibase
  10.1140/epjc/s10052-015-3720-2} {\bibfield  {journal} {\bibinfo  {journal}
  {Eur. Phys. J. C}\ }\textbf {\bibinfo {volume} {75}},\ \bibinfo {pages} {495}
  (\bibinfo {year} {2015})},\ \Eprint {http://arxiv.org/abs/1508.00651}
  {arXiv:1508.00651 [hep-ph]} \BibitemShut {NoStop}%
\bibitem [{\citenamefont {Zhang}\ \emph {et~al.}(2017)\citenamefont {Zhang},
  \citenamefont {Hou}, \citenamefont {Kojo},\ and\ \citenamefont
  {Qin}}]{Zhang:2017icm}%
  \BibitemOpen
  \bibfield  {author} {\bibinfo {author} {\bibfnamefont {H.}~\bibnamefont
  {Zhang}}, \bibinfo {author} {\bibfnamefont {D.}~\bibnamefont {Hou}}, \bibinfo
  {author} {\bibfnamefont {T.}~\bibnamefont {Kojo}}, \ and\ \bibinfo {author}
  {\bibfnamefont {B.}~\bibnamefont {Qin}},\ }\href {\doibase
  10.1103/PhysRevD.96.114029} {\bibfield  {journal} {\bibinfo  {journal} {Phys.
  Rev. D}\ }\textbf {\bibinfo {volume} {96}},\ \bibinfo {pages} {114029}
  (\bibinfo {year} {2017})},\ \Eprint {http://arxiv.org/abs/1709.05654}
  {arXiv:1709.05654 [hep-ph]} \BibitemShut {NoStop}%
\bibitem [{\citenamefont {Bazavov}\ \emph {et~al.}(2017)\citenamefont {Bazavov}
  \emph {et~al.}}]{Bazavov:2017tot}%
  \BibitemOpen
  \bibfield  {author} {\bibinfo {author} {\bibfnamefont {A.}~\bibnamefont
  {Bazavov}} \emph {et~al.} (\bibinfo {collaboration} {HotQCD}),\ }\href
  {\doibase 10.1103/PhysRevD.96.074510} {\bibfield  {journal} {\bibinfo
  {journal} {Phys. Rev. D}\ }\textbf {\bibinfo {volume} {96}},\ \bibinfo
  {pages} {074510} (\bibinfo {year} {2017})},\ \Eprint
  {http://arxiv.org/abs/1708.04897} {arXiv:1708.04897 [hep-lat]} \BibitemShut
  {NoStop}%
\bibitem [{\citenamefont {Fu}\ \emph {et~al.}(2020)\citenamefont {Fu},
  \citenamefont {Pawlowski},\ and\ \citenamefont {Rennecke}}]{Fu:2019hdw}%
  \BibitemOpen
  \bibfield  {author} {\bibinfo {author} {\bibfnamefont {W.-j.}\ \bibnamefont
  {Fu}}, \bibinfo {author} {\bibfnamefont {J.~M.}\ \bibnamefont {Pawlowski}}, \
  and\ \bibinfo {author} {\bibfnamefont {F.}~\bibnamefont {Rennecke}},\ }\href
  {\doibase 10.1103/PhysRevD.101.054032} {\bibfield  {journal} {\bibinfo
  {journal} {Phys. Rev. D}\ }\textbf {\bibinfo {volume} {101}},\ \bibinfo
  {pages} {054032} (\bibinfo {year} {2020})},\ \Eprint
  {http://arxiv.org/abs/1909.02991} {arXiv:1909.02991 [hep-ph]} \BibitemShut
  {NoStop}%
\bibitem [{\citenamefont {Fischer}(2019)}]{Fischer:2018sdj}%
  \BibitemOpen
  \bibfield  {author} {\bibinfo {author} {\bibfnamefont {C.~S.}\ \bibnamefont
  {Fischer}},\ }\href {\doibase 10.1016/j.ppnp.2019.01.002} {\bibfield
  {journal} {\bibinfo  {journal} {Prog. Part. Nucl. Phys.}\ }\textbf {\bibinfo
  {volume} {105}},\ \bibinfo {pages} {1} (\bibinfo {year} {2019})},\ \Eprint
  {http://arxiv.org/abs/1810.12938} {arXiv:1810.12938 [hep-ph]} \BibitemShut
  {NoStop}%
\bibitem [{\citenamefont {Caines}(2017)}]{Caines:2017search}%
  \BibitemOpen
  \bibfield  {author} {\bibinfo {author} {\bibfnamefont {H.}~\bibnamefont
  {Caines}},\ }\href@noop {} {\bibfield  {journal} {\bibinfo  {journal}
  {Nuclear Physics A}\ }\textbf {\bibinfo {volume} {967}},\ \bibinfo {pages}
  {121} (\bibinfo {year} {2017})}\BibitemShut {NoStop}%
\bibitem [{\citenamefont {Stephanov}\ \emph {et~al.}(1998)\citenamefont
  {Stephanov}, \citenamefont {Rajagopal},\ and\ \citenamefont
  {Shuryak}}]{Stephanov:1998dy}%
  \BibitemOpen
  \bibfield  {author} {\bibinfo {author} {\bibfnamefont {M.~A.}\ \bibnamefont
  {Stephanov}}, \bibinfo {author} {\bibfnamefont {K.}~\bibnamefont
  {Rajagopal}}, \ and\ \bibinfo {author} {\bibfnamefont {E.~V.}\ \bibnamefont
  {Shuryak}},\ }\href {\doibase 10.1103/PhysRevLett.81.4816} {\bibfield
  {journal} {\bibinfo  {journal} {Phys. Rev. Lett.}\ }\textbf {\bibinfo
  {volume} {81}},\ \bibinfo {pages} {4816} (\bibinfo {year} {1998})},\ \Eprint
  {http://arxiv.org/abs/hep-ph/9806219} {arXiv:hep-ph/9806219} \BibitemShut
  {NoStop}%
\bibitem [{\citenamefont {Bzdak}\ \emph {et~al.}(2020)\citenamefont {Bzdak},
  \citenamefont {Esumi}, \citenamefont {Koch}, \citenamefont {Liao},
  \citenamefont {Stephanov},\ and\ \citenamefont {Xu}}]{Bzdak:2019pkr}%
  \BibitemOpen
  \bibfield  {author} {\bibinfo {author} {\bibfnamefont {A.}~\bibnamefont
  {Bzdak}}, \bibinfo {author} {\bibfnamefont {S.}~\bibnamefont {Esumi}},
  \bibinfo {author} {\bibfnamefont {V.}~\bibnamefont {Koch}}, \bibinfo {author}
  {\bibfnamefont {J.}~\bibnamefont {Liao}}, \bibinfo {author} {\bibfnamefont
  {M.}~\bibnamefont {Stephanov}}, \ and\ \bibinfo {author} {\bibfnamefont
  {N.}~\bibnamefont {Xu}},\ }\href {\doibase 10.1016/j.physrep.2020.01.005}
  {\bibfield  {journal} {\bibinfo  {journal} {Phys. Rept.}\ }\textbf {\bibinfo
  {volume} {853}},\ \bibinfo {pages} {1} (\bibinfo {year} {2020})},\ \Eprint
  {http://arxiv.org/abs/1906.00936} {arXiv:1906.00936 [nucl-th]} \BibitemShut
  {NoStop}%
\bibitem [{\citenamefont {Adamczyk}\ \emph {et~al.}(2014)\citenamefont
  {Adamczyk} \emph {et~al.}}]{Adamczyk:2013dal}%
  \BibitemOpen
  \bibfield  {author} {\bibinfo {author} {\bibfnamefont {L.}~\bibnamefont
  {Adamczyk}} \emph {et~al.} (\bibinfo {collaboration} {STAR}),\ }\href
  {\doibase 10.1103/PhysRevLett.112.032302} {\bibfield  {journal} {\bibinfo
  {journal} {Phys. Rev. Lett.}\ }\textbf {\bibinfo {volume} {112}},\ \bibinfo
  {pages} {032302} (\bibinfo {year} {2014})},\ \Eprint
  {http://arxiv.org/abs/1309.5681} {arXiv:1309.5681 [nucl-ex]} \BibitemShut
  {NoStop}%
\bibitem [{\citenamefont {Adam}\ \emph {et~al.}(2020)\citenamefont {Adam},
  \citenamefont {Adamczyk}, \citenamefont {Adams}, \citenamefont {Adkins},
  \citenamefont {Agakishiev}, \citenamefont {Aggarwal}, \citenamefont
  {Ahammed}, \citenamefont {Alekseev}, \citenamefont {Anderson}, \citenamefont
  {Aparin} \emph {et~al.}}]{Adam:2020unf}%
  \BibitemOpen
  \bibfield  {author} {\bibinfo {author} {\bibfnamefont {J.}~\bibnamefont
  {Adam}}, \bibinfo {author} {\bibfnamefont {L.}~\bibnamefont {Adamczyk}},
  \bibinfo {author} {\bibfnamefont {J.}~\bibnamefont {Adams}}, \bibinfo
  {author} {\bibfnamefont {J.}~\bibnamefont {Adkins}}, \bibinfo {author}
  {\bibfnamefont {G.}~\bibnamefont {Agakishiev}}, \bibinfo {author}
  {\bibfnamefont {M.}~\bibnamefont {Aggarwal}}, \bibinfo {author}
  {\bibfnamefont {Z.}~\bibnamefont {Ahammed}}, \bibinfo {author} {\bibfnamefont
  {I.}~\bibnamefont {Alekseev}}, \bibinfo {author} {\bibfnamefont
  {D.}~\bibnamefont {Anderson}}, \bibinfo {author} {\bibfnamefont
  {A.}~\bibnamefont {Aparin}},  \emph {et~al.},\ }\href@noop {} {\bibfield
  {journal} {\bibinfo  {journal} {arXiv preprint arXiv:2001.02852}\ } (\bibinfo
  {year} {2020})}\BibitemShut {NoStop}%
\bibitem [{\citenamefont {Adams}\ \emph {et~al.}(2020)\citenamefont {Adams}
  \emph {et~al.}}]{Adams:2019fpo}%
  \BibitemOpen
  \bibfield  {author} {\bibinfo {author} {\bibfnamefont {J.}~\bibnamefont
  {Adams}} \emph {et~al.},\ }\href {\doibase 10.1016/j.nima.2020.163970}
  {\bibfield  {journal} {\bibinfo  {journal} {Nucl. Instrum. Meth. A}\ }\textbf
  {\bibinfo {volume} {968}},\ \bibinfo {pages} {163970} (\bibinfo {year}
  {2020})},\ \Eprint {http://arxiv.org/abs/1912.05243} {arXiv:1912.05243
  [physics.ins-det]} \BibitemShut {NoStop}%
\bibitem [{\citenamefont {Bass}\ \emph {et~al.}(1998)\citenamefont {Bass} \emph
  {et~al.}}]{Bass:1998ca}%
  \BibitemOpen
  \bibfield  {author} {\bibinfo {author} {\bibfnamefont {S.}~\bibnamefont
  {Bass}} \emph {et~al.},\ }\href {\doibase 10.1016/S0146-6410(98)00058-1}
  {\bibfield  {journal} {\bibinfo  {journal} {Prog. Part. Nucl. Phys.}\
  }\textbf {\bibinfo {volume} {41}},\ \bibinfo {pages} {255} (\bibinfo {year}
  {1998})},\ \Eprint {http://arxiv.org/abs/nucl-th/9803035}
  {arXiv:nucl-th/9803035} \BibitemShut {NoStop}%
\bibitem [{\citenamefont {Miller}\ \emph {et~al.}(2007)\citenamefont {Miller},
  \citenamefont {Reygers}, \citenamefont {Sanders},\ and\ \citenamefont
  {Steinberg}}]{Miller:2007ri}%
  \BibitemOpen
  \bibfield  {author} {\bibinfo {author} {\bibfnamefont {M.~L.}\ \bibnamefont
  {Miller}}, \bibinfo {author} {\bibfnamefont {K.}~\bibnamefont {Reygers}},
  \bibinfo {author} {\bibfnamefont {S.~J.}\ \bibnamefont {Sanders}}, \ and\
  \bibinfo {author} {\bibfnamefont {P.}~\bibnamefont {Steinberg}},\ }\href
  {\doibase 10.1146/annurev.nucl.57.090506.123020} {\bibfield  {journal}
  {\bibinfo  {journal} {Ann. Rev. Nucl. Part. Sci.}\ }\textbf {\bibinfo
  {volume} {57}},\ \bibinfo {pages} {205} (\bibinfo {year} {2007})},\ \Eprint
  {http://arxiv.org/abs/nucl-ex/0701025} {arXiv:nucl-ex/0701025} \BibitemShut
  {NoStop}%
\bibitem [{\citenamefont {Bleicher}\ \emph {et~al.}(1999)\citenamefont
  {Bleicher} \emph {et~al.}}]{Bleicher:1999xi}%
  \BibitemOpen
  \bibfield  {author} {\bibinfo {author} {\bibfnamefont {M.}~\bibnamefont
  {Bleicher}} \emph {et~al.},\ }\href {\doibase 10.1088/0954-3899/25/9/308}
  {\bibfield  {journal} {\bibinfo  {journal} {J. Phys. G}\ }\textbf {\bibinfo
  {volume} {25}},\ \bibinfo {pages} {1859} (\bibinfo {year} {1999})},\ \Eprint
  {http://arxiv.org/abs/hep-ph/9909407} {arXiv:hep-ph/9909407} \BibitemShut
  {NoStop}%
\bibitem [{\citenamefont {Adamczyk}\ \emph {et~al.}(2015)\citenamefont
  {Adamczyk} \emph {et~al.}}]{Adamczyk:2015eqo}%
  \BibitemOpen
  \bibfield  {author} {\bibinfo {author} {\bibfnamefont {L.}~\bibnamefont
  {Adamczyk}} \emph {et~al.} (\bibinfo {collaboration} {STAR}),\ }\href
  {\doibase 10.1103/PhysRevLett.114.252302} {\bibfield  {journal} {\bibinfo
  {journal} {Phys. Rev. Lett.}\ }\textbf {\bibinfo {volume} {114}},\ \bibinfo
  {pages} {252302} (\bibinfo {year} {2015})},\ \Eprint
  {http://arxiv.org/abs/1504.02175} {arXiv:1504.02175 [nucl-ex]} \BibitemShut
  {NoStop}%
\bibitem [{\citenamefont {Adamczyk}\ \emph {et~al.}(2013)\citenamefont
  {Adamczyk} \emph {et~al.}}]{Adamczyk:2013hsi}%
  \BibitemOpen
  \bibfield  {author} {\bibinfo {author} {\bibfnamefont {L.}~\bibnamefont
  {Adamczyk}} \emph {et~al.} (\bibinfo {collaboration} {STAR}),\ }\href
  {\doibase 10.1103/PhysRevC.88.064911} {\bibfield  {journal} {\bibinfo
  {journal} {Phys. Rev. C}\ }\textbf {\bibinfo {volume} {88}},\ \bibinfo
  {pages} {064911} (\bibinfo {year} {2013})},\ \Eprint
  {http://arxiv.org/abs/1302.3802} {arXiv:1302.3802 [nucl-ex]} \BibitemShut
  {NoStop}%
\bibitem [{\citenamefont {Anderson}\ \emph {et~al.}(2003)\citenamefont
  {Anderson}, \citenamefont {Berkovitz}, \citenamefont {Betts}, \citenamefont
  {Bossingham}, \citenamefont {Bieser}, \citenamefont {Brown}, \citenamefont
  {Burks}, \citenamefont {de~la Barca~S{\'a}nchez}, \citenamefont {Cebra},
  \citenamefont {Cherney} \emph {et~al.}}]{anderson2003star}%
  \BibitemOpen
  \bibfield  {author} {\bibinfo {author} {\bibfnamefont {M.}~\bibnamefont
  {Anderson}}, \bibinfo {author} {\bibfnamefont {J.}~\bibnamefont {Berkovitz}},
  \bibinfo {author} {\bibfnamefont {W.}~\bibnamefont {Betts}}, \bibinfo
  {author} {\bibfnamefont {R.}~\bibnamefont {Bossingham}}, \bibinfo {author}
  {\bibfnamefont {F.}~\bibnamefont {Bieser}}, \bibinfo {author} {\bibfnamefont
  {R.}~\bibnamefont {Brown}}, \bibinfo {author} {\bibfnamefont
  {M.}~\bibnamefont {Burks}}, \bibinfo {author} {\bibfnamefont {M.~C.}\
  \bibnamefont {de~la Barca~S{\'a}nchez}}, \bibinfo {author} {\bibfnamefont
  {D.}~\bibnamefont {Cebra}}, \bibinfo {author} {\bibfnamefont
  {M.}~\bibnamefont {Cherney}},  \emph {et~al.},\ }\href@noop {} {\bibfield
  {journal} {\bibinfo  {journal} {Nuclear Instruments and Methods in Physics
  Research Section A: Accelerators, Spectrometers, Detectors and Associated
  Equipment}\ }\textbf {\bibinfo {volume} {499}},\ \bibinfo {pages} {659}
  (\bibinfo {year} {2003})}\BibitemShut {NoStop}%
\bibitem [{Note1()}]{Note1}%
  \BibitemOpen
  \bibinfo {note} {Our simulation does not include the effects of the magnetic
  field in the STAR experiment; however, at these rapidities, this is
  relatively unimportant for measuring multiplicity as a function of $\eta $.
  Secondary scattering and production in surrounding material (magnet iron,
  beam pipe, etc) is a larger effect that is not accounted for, in our
  simulation. Exploring corrections due to these effects is outside the scope
  of the present study.}\BibitemShut {Stop}%
\bibitem [{\citenamefont {Luo}\ \emph {et~al.}(2013)\citenamefont {Luo},
  \citenamefont {Xu}, \citenamefont {Mohanty},\ and\ \citenamefont
  {Xu}}]{Luo:2013bmi}%
  \BibitemOpen
  \bibfield  {author} {\bibinfo {author} {\bibfnamefont {X.}~\bibnamefont
  {Luo}}, \bibinfo {author} {\bibfnamefont {J.}~\bibnamefont {Xu}}, \bibinfo
  {author} {\bibfnamefont {B.}~\bibnamefont {Mohanty}}, \ and\ \bibinfo
  {author} {\bibfnamefont {N.}~\bibnamefont {Xu}},\ }\href {\doibase
  10.1088/0954-3899/40/10/105104} {\bibfield  {journal} {\bibinfo  {journal}
  {J. Phys. G}\ }\textbf {\bibinfo {volume} {40}},\ \bibinfo {pages} {105104}
  (\bibinfo {year} {2013})},\ \Eprint {http://arxiv.org/abs/1302.2332}
  {arXiv:1302.2332 [nucl-ex]} \BibitemShut {NoStop}%
\end{thebibliography}%

\end{document}